\documentclass[a4paper,12pt,authoryear]{article}
\usepackage{xcolor,amsmath}
\usepackage{amsthm,bm}
\usepackage{hyperref}
\usepackage{amssymb}
\usepackage{dsfont}
\usepackage{comment}
\usepackage{float}
\usepackage{graphicx}
\usepackage{slashbox}
\usepackage{pbox}
\usepackage{colortbl}
\usepackage{epstopdf}
\usepackage{mathtools}
\usepackage{multirow}
\usepackage{relsize}
\usepackage{bigints}
\usepackage{cases}
\usepackage[font={footnotesize}]{caption}
\usepackage{hhline}
\usepackage{booktabs}

\usepackage{natbib}

\newtheorem{theorem}{Theorem}

\theoremstyle{remark}

\theoremstyle{plain}

\newcommand\myiid{\mathrel{\overset{\makebox[0pt]{\mbox{\normalfont\tiny\sffamily i.i.d.}}}{\sim}}}

\newcommand*\diff{\mathop{}\!\mathrm{d}}

\title{Bayesian Loss-based Approach to Change Point Analysis}

\usepackage{authblk}

\author[1]{Laurentiu Hinoveanu\thanks{lch36@kent.ac.uk}}
\author[1]{Fabrizio Leisen\thanks{F.Leisen@kent.ac.uk}}
\author[1]{Cristiano Villa\thanks{C.Villa-88@kent.ac.uk}}
\affil[1]{School of Mathematics, Statistics and Actuarial Science, University of Kent}

\usepackage[english]{isodate}

\usepackage{parskip}
\date{}

\begin{document}
\maketitle

\begin{abstract}
\noindent
In this paper we present a loss-based approach to change point analysis. In particular, we look at the problem from two perspectives. The first focuses on the definition of a prior when the number of change points is known a priori. The second contribution aims to estimate the number of change points by using a loss-based approach recently introduced in the literature. The latter considers change point estimation as a model selection exercise. We show the performance of the proposed approach on simulated data and real data sets.
\end{abstract}
\textbf{Keywords:} Change point; Discrete parameter space; Loss-based prior; Model selection.

\section{Introduction}\label{sc_introduction}

There are several practical scenarios where it is inappropriate to assume that the distribution of the observations does not change. For example, financial data sets can exhibit alternate behaviours due to crisis periods. In this case it is sensible to assume changes in the underlying distribution. The change in the distribution can be either in the value of one or more of the parameters or, more in general, on the family of the distribution. In the latter case, for example, one may deem appropriate to consider a normal density for the stagnation periods, while a Student $t$, with relatively heavy tails, may be more suitable to represent observations in the more turbulent stages of a crisis. The task of identifying if, and when, one or more changes have occurred is not trivial and requires appropriate methods to avoid detection of a large number of changes or, at the opposite extreme, seeing no changes at all. 
The change point problem has been deeply studied from a Bayesian point of view. \citet{Chernoff1964} focused on the change in the means of normally distributed variables. \citet{Smith1975} looked into the single change point problem when different knowledge of the parameters of the underlying distributions is available: all known, some of them known or none of them known. \citet{Smith1975} focuses on the binomial and normal distributions. In \citet{Muliere1985} the problem is tackled from a Bayesian nonparametric perspective. The authors consider Dirichlet processes with independent base measures as underlying distributions. In this framework, \citet{Petrone1997} have showed that the Dirichlet process prior could have a strong effect on the inference and may lead to wrong conclusions in the case of a single change point. \citet{Raftery} have approached the single change point problem in the context of a Poisson likelihood under both proper and improper priors for the model parameters. \citet{Carlin} build on the work of \citet{Raftery} by considering a two level hierarchical model.  Both papers illustrate the respective approaches by studying the well-known British coal-mining disaster data set. In the context of multiple change points detection, \citet{Loschi2005} have provided a fully Bayesian treatment for the product partitions model of \citet{Barry1992}. Their application focused on stock exchange data. \citet{Stephens1994} has extended the Gibbs sampler introduced by \citet{Carlin} in the change point literature to handle multiple change points. \citet{Hannart2009} have used Bayesian decision theory, in particular 0-1 cost functions, to estimate multiple changes in homoskedastic normally distributed observations. \citet{Schwaller2017} extend the product partition model of \citet{Barry1992} by adding a graphical structure which could capture the dependencies between multivariate observations. \citet{Fearnhead2007} proposed a filtering algorithm for the sequential  multiple change points detection problem in the case of piecewise regression models.
\citet{Henderson1993} introduced a partial Bayesian approach which involves the use of a profile likelihood, where the aim is to detect multiple changes in the mean of Poisson distributions with an application to haemolytic uraemic syndrome (HUS) data. The same data set was studied by \citet{Tian2009}, who proposed a method which treats the change points as latent variables. 
\citet{ko2015} have proposed an extension to the hidden Markov model of \citet{Chib} by using a Dirichlet process prior on each row of the regime matrix. Their model is semiparametric,  as the number of states is not specified in advance, but it grows according to the data size. \citet{Heard2017} have proposed a new sequential Monte Carlo algorithm to infer multiple change points.

Whilst the literature covering change point analysis from a Bayesian perspective is vast when prior distributions are elicited, the documentation referring to analysis under minimal prior information is limited, see \citet{Moreno2005} and \cite{Giron}. The former paper discusses the single change point problem in a model selection setting, whilst the latter paper, which is an extension of the former, tackles the multivariate change point problem in the context of linear regression models. Our work aims to contribute to the methodology for change point analysis under the assumption that the information about the number of change points and their location is minimal. First, we discuss the definition of an objective prior for change point location, both for single and multiple  changes, assuming the number of changes is known a priori. Then, we define a prior on the number of change points via a model selection approach. Here, we assume that the change point coincides with one of the observations. As such, given $X_1, X_2,\dots,X_n$ data points, the change point location is discrete. To the best of our knowledge, the sole general objective approach to define prior distributions on discrete spaces is the one introduced by \citet{VillaMass}. 

To illustrate the idea, consider a probability distribution $f(x|m)$, where $m\in \mathbb{M}$ is a discrete parameter. Then, the prior $\pi(m)$ is obtained by objectively measuring what is lost if the value $m$ is removed from the parameter space, and it is the true value. According to \citet{berk1966}, if a model is misspecified, the posterior distribution  asymptotically accumulates on the model which is the most similar to the true one, where the similarity is measured in terms of the Kullback--Leibler (KL) divergence. Therefore, $D_{KL}(f(\cdot|m)\|f(\cdot|m^{\prime}))$, where $m^{\prime}$ is the parameter characterising the nearest model to $f(x|m)$, represents the utility of keeping $m$. The objective prior is then obtained by linking the aforementioned utility via the self-information loss:
\begin{align}
\pi(m)& \propto  \exp\left\lbrace\displaystyle\min_{m^{\prime} \neq m }D_{KL}(f(\cdot|m)\|f(\cdot|m^{\prime}))\right\rbrace-1,
\label{VillaMass}
\end{align} 
where the Kullback--Leibler divergence  \citep{kullback1951} from the sampling distribution with density $f(x|m)$ to the one with density $f(x|m^{\prime})$ is defined as:
\begin{equation*}
D_{KL}(f(\cdot|m)\|f(\cdot|m^{\prime}))=
      \displaystyle\mathlarger{\int}\displaylimits_{\mathcal{X}} f(x|m) \cdot \log\left[\dfrac{f(x|m)}{f(x|m^{\prime})}\right]\diff x.
   \label{KLdef}
\end{equation*}
Throughout the paper, the objective prior defined in equation \eqref{VillaMass} will be referenced as the \textit{loss-based} prior. This approach is used to define an objective prior distribution when the number of change points is known a priori. To obtain a prior distribution for the number of change points, we adopt a model selection approach based on the results in \citet{VillaModel}, where a method to define a prior on the space of models is proposed.
To illustrate, let us consider $k$ Bayesian models:
\begin{equation}
M_{j}=\{f_{j}(x|\theta_{j}), \pi_{j}(\theta_{j})\}\qquad   j\in\{1,2,\ldots, k\},
\label{modeldef}
\end{equation}

where $f_{j}(x|\theta_{j})$ is the sampling density characterised by $\theta_{j}$ and  $\pi_{j}(\theta_{j})$ represents the prior on the model parameter.

Assuming the prior on the model parameter, $\pi_{j}(\theta_{j})$, is proper, the model prior probability $\Pr(M_{j})$ is proportional to the expected minimum Kullback--Leibler divergence from $M_j$, where the expectation is considered with respect to  $\pi_{j}(\theta_{j})$. That is: 
\begin{align}
\Pr(M_{j})\propto &\exp\left\lbrace  \mathbb{E}_{\pi_{j}}\left[\inf_{\theta_{i},i\neq j} D_{KL}(f_{j}(x|\theta_{j})\|f_{i}(x|\theta_{i}))\right]\right\rbrace \qquad    j=1, \ldots, k. \label{VillaModelCompactOrg}
\end{align}

%Across this paper, we use the following shorthand notation for equation \eqref{VillaModelCompactOrg}:
%\begin{align*}
%\Pr(M_{j})\propto &\exp\left\lbrace  \mathbb{E}_{\pi_{j}}\left[\inf_{\theta_{i},i\neq j} D_{KL}(M_j\|M_i)\right]\right\rbrace\qquad    j=1, \ldots, k. 
%\end{align*}

The model prior probabilities defined in equation \eqref{VillaModelCompactOrg} can be employed to derive the model posterior probabilities through:
\begin{equation}
\Pr(M_{i}|x)= \left[\sum_{j=1}^{k}\dfrac{\Pr(M_{j})}{\Pr(M_{i})} B_{ji}\right]^{-1},
\label{modelposterior}
\end{equation}
where $B_{ji}$ is the Bayes factor between model $M_{j}$ and model $M_{i}$, defined as 
\begin{equation*}
B_{ji}=\dfrac{\int f_{j}(x|\theta_{j})\pi_{j}(\theta_{j})\diff \theta_{j}}{\int f_{i}(x|\theta_{i})\pi_{i}(\theta_{i})\diff \theta_{i}},
\end{equation*}
with $i\neq j \in \{1, 2, \ldots, k\}$. 

This paper is structured as follows: in Section \ref{sc_locations} we establish the way we set objective priors on both single and multiple change point locations. Section \ref{sc_numberCP} shows how we define the model prior probabilities for the number of change point locations. Illustrations of the model selection exercise are provided in Sections \ref{sc_simulation} and \ref{sc_realdata}, where we work with simulated and real data, respectively. Section \ref{Conclusion} is dedicated to final remarks.

\section{Objective Prior on the Change Point Locations}\label{sc_locations}
This section is devoted to the derivation of the loss-based prior when the number of change points is known a priori. Specifically, let $k$ be the number of change points and $m_1<m_2<\ldots<m_k$ their locations. We introduce the idea in the simple case where we assume that there is only one change point in the data set (see Section \ref{sub_sc_OneChgPoint}). Then, we extend the results to the more general case where multiple change points are assumed (see Section \ref{sub_sc_MultipleChgPoints}). 

A well-known objective prior for finite parameter spaces, in cases where there is no structure, is the uniform prior \citep{BBS2012}. As such, a natural choice for the prior on the change points location is the uniform \citep{Koop}. The corresponding loss-based prior is indeed the uniform, as shown below, which is a reassuring result as the objective prior for a specific parameter space, if exists, should be unique.

\subsection{Single Change Point}
\label{sub_sc_OneChgPoint}
As mentioned above, we show that the loss-based prior for the single change point case coincides with the discrete uniform distribution over the set $\{1, 2, \ldots, n-1\}$.

Let $\mathbf{X}^{(n)}=(X_{1},\ldots, X_{n})$ denote an \textit{n}-dimensional vector of random variables, representing the random sample, and $m$ be our single change point location, that is $m \in \{1, 2, \ldots, n-1\}$, such that
\begin{align}
X_{1}, \ldots, X_{m}| \tilde\theta_{1} &\myiid f_{1}(\cdot|\tilde\theta_{1}) \nonumber \\
X_{m+1}, \ldots, X_{n}| \tilde\theta_{2} &\myiid f_{2}(\cdot|\tilde\theta_{2}).
\label{SingleChangePointCase}
\end{align}
Note that we assume that there is a change point in the series, as such the space of $m$ does not include the case $m=n$. In addition, we assume that $\tilde\theta_1 \neq \tilde\theta_2$ when $f_1=f_2$. The sampling density for the vector of observations  $\mathbf{x}^{(n)}=(x_1,\dots,x_n)$ is:
\begin{equation}
f(\mathbf{x}^{(n)}|m,\tilde\theta_{1},\tilde\theta_{2})=\prod_{i=1}^{m} f_{1}(x_{i}|\tilde\theta_{1})\prod_{i=m+1}^{n} f_{2}(x_{i}|\tilde\theta_{2}).
\label{likelihood}
\end{equation}
Let $m^{\prime} \neq m$. Then, the Kullback--Leibler divergence between the model parametrised by $m$ and the one parametrised by $m^{\prime}$ is:
\begin{align}
D_{KL}(f(\mathbf{x}^{(n)}|m,\tilde\theta_{1},\tilde\theta_{2})\|f(\mathbf{x}^{(n)}|m^{\prime},\tilde\theta_{1},\tilde\theta_{2}))=& \int f(\mathbf{x}^{(n)}|m,\tilde\theta_{1},\tilde\theta_{2})  \nonumber \\ &\hspace*{1em}\log\left(\dfrac{f(\mathbf{x}^{(n)}|m,\tilde\theta_{1},\tilde\theta_{2})}{f(\mathbf{x}^{(n)}|m^{\prime},\tilde\theta_{1},\tilde\theta_{2})}\right)\,\mathrm{d}\mathbf{x}^{(n)}.
\label{klremarkOneChg}
\end{align}
Without loss of generality, consider $m<m^{\prime}$. In this case, note that
\begin{align*}
\dfrac{f(\mathbf{x}^{(n)}|m,\tilde\theta_{1},\tilde\theta_{2})}{f(\mathbf{x}^{(n)}|m^{\prime},\tilde\theta_{1},\tilde\theta_{2})}=\prod_{i=m+1}^{m^{\prime}} \dfrac{f_{2}(x_{i}|\tilde\theta_{2})}{f_{1}(x_{i}|\tilde\theta_{1})},
\end{align*}
leading to
%\begin{align}
%&D_{KL}(f(\mathbf{x}^{(n)}|m,\tilde\theta_{1},\tilde\theta_{2})\|f(\mathbf{x}^{(n)}|m^{\prime},\tilde\theta_{1},\tilde\theta_{2}))=\int \left( \prod_{j=1}^{m} f_{1}(x_{j}|\tilde\theta_{1})\right)  \nonumber\\ &\hspace*{1em}\cdot \left( \prod_{i=m+1}^{m^{\prime}} f_{2}(x_{i}|\tilde\theta_{2})\right) \cdot \left( \prod_{j=m^{\prime}+1}^{n} f_{2}(x_{j}|\tilde\theta_{2})\right) \cdot \left[\sum_{i=m+1}^{m^{\prime}} \ln\left(\dfrac{f_{2}(x_{i}|\tilde\theta_{2})}{f_{1}(x_{i}|\tilde\theta_{1})}\right)\right]\,\mathrm{d}\mathbf{x}^{(n)} \nonumber \\
%&=\sum_{i=m+1}^{m^{\prime}} \int \left( \prod_{j=1}^{m} f_{1}(x_{j}|\tilde\theta_{1})\right) \cdot \left( \prod_{j=m^{\prime}+1}^{n} f_{2}(x_{j}|\tilde\theta_{2})\right)\cdot \left(\prod_{\substack{r=m+1\\ r \neq i}}^{m^{\prime}} f_{2}(x_{r}|\tilde\theta_{2})\right) \nonumber  \\ &\hspace*{1em}\cdot \left[ f_{2}(x_{i}|\tilde\theta_{2})\cdot\ln\left(\dfrac{f_{2}(x_{i}|\tilde\theta_{2})}{f_{1}(x_{i}|\tilde\theta_{1})}\right)\right]\,\mathrm{d}\mathbf{x}^{(n)}.
%\label{step1}
%\end{align}
%Integrating with respect to all variables which are not indexed by $i$ in equation \eqref{step1}, we obtain: %\cdot \phantom{\dfrac{1}{2}}\right. \nonumber\\ &\left.
\begin{align}
D_{KL}(f(\mathbf{x}^{(n)}|m,\tilde\theta_{1},\tilde\theta_{2})&\|f(\mathbf{x}^{(n)}|m^{\prime},\tilde\theta_{1},\tilde\theta_{2}))=\nonumber \\ &\sum_{i=m+1}^{m^{\prime}}  \int f_{2}(x_{i}|\tilde\theta_{2})\log\left(\dfrac{f_{2}(x_{i}|\tilde\theta_{2})}{f_{1}(x_{i}|\tilde\theta_{1})}\right)\,\mathrm{d}x_{i}.
\label{step2}
\end{align}
On the right hand side of equation \eqref{step2}, we can recognise the Kullback--Leibler divergence from density $f_{2}$ to density $f_{1}$, thus getting:
\begin{align}
D_{KL}(f(\mathbf{x}^{(n)}|m,\tilde\theta_{1},\tilde\theta_{2})&||f(\mathbf{x}^{(n)}|m^{\prime},\tilde\theta_{1},\tilde\theta_{2}))=\nonumber \\ 
&\hspace{2cm}(m^{\prime}-m) D_{KL}(f_{2}(\cdot|\tilde\theta_{2})\|f_{1}(\cdot|\tilde\theta_{1})).
\label{step3case1}
\end{align}
%By analogy with the steps taken in Equations \eqref{step1},\eqref{step2} and \eqref{step3case1}, we derive the following relation for the case $m>m^{\prime}$:
In a similar fashion, when $m>m^{\prime}$, we have that:
\begin{align}
D_{KL}(f(\mathbf{x}^{(n)}|m,\tilde\theta_{1},\tilde\theta_{2})&\|f(\mathbf{x}^{(n)}|m^{\prime},\tilde\theta_{1},\tilde\theta_{2}))=\nonumber \\ 
&\hspace{2cm}(m-m^{\prime}) D_{KL}(f_{1}(\cdot|\tilde\theta_{1})\|f_{2}(\cdot|\tilde\theta_{2})).
\label{step3case2}
\end{align}

In this single change point scenario, we can consider $m^{\prime}$ as a perturbation of the change point location $m$, that is $m^{\prime}=m\pm l$ where $l \in \mathbb{N}^{*}$, such that $1\leq m^{\prime}<n$. Then, taking into account equations \eqref{step3case1} and \eqref{step3case2}, the Kullback--Leibler divergence becomes:

\begin{align*}
D_{KL}(f(\mathbf{x}^{(n)}|m,\tilde\theta_{1},\tilde\theta_{2})\|f(\mathbf{x}^{(n)}&|m^{\prime},\tilde\theta_{1},\tilde\theta_{2}))=\\
&\left\{
\begin{array}{ll}
l \cdot D_{KL}(f_{2}(\cdot|\tilde\theta_{2})\|f_{1}(\cdot|\tilde\theta_{1})), & \mbox{ if } m<m^{\prime}\\
&\\
l \cdot D_{KL}(f_{1}(\cdot|\tilde\theta_{1})\|f_{2}(\cdot|\tilde\theta_{2})), & \mbox{ if } m>m^{\prime},
\end{array}\right.\\
\end{align*}
and
\begin{align}
\min_{m^{\prime}\neq m}&\left[D_{KL}(f(\mathbf{x^{(n)}}|m,\tilde\theta_{1},\tilde\theta_{2})\|f(\mathbf{x^{(n)}}|m^{\prime},\tilde\theta_{1},\tilde\theta_{2}))\right]=\nonumber \\&=\min_{m^{\prime}\neq m}\lbrace l \cdot D_{KL}(f_{2}(\cdot|\tilde\theta_{2})\|f_{1}(\cdot|\tilde\theta_{1})),l \cdot D_{KL}(f_{1}(\cdot|\tilde\theta_{1})\|f_{2}(\cdot|\tilde\theta_{2}))\rbrace\nonumber \\&= \min_{m^{\prime}\neq m}\lbrace D_{KL}(f_{2}(\cdot|\tilde\theta_{2})\|f_{1}(\cdot|\tilde\theta_{1})),D_{KL}(f_{1}(\cdot|\tilde\theta_{1})\|f_{2}(\cdot|\tilde\theta_{2}))\rbrace  \cdot \underbrace{\min_{m^{\prime}\neq m}\{l\}}_{1}.
\label{KL_no_m}
\end{align}
%From equation \eqref{KL_no_m}, we observe that $\displaystyle\min_{m^{\prime}\neq m}\left[D_{KL}(f(\mathbf{x^{(n)}}|m,\tilde\theta_{1},\tilde\theta_{2})\|\right.\\ \left.f(\mathbf{x^{(n)}}|m^{\prime},\tilde\theta_{1},\tilde\theta_{2}))\right]$ is only a function of $\tilde\theta_{1}$ and $\tilde\theta_{2}$ and does not depend by $m$. Thus, if we apply the Loss-based prior from equation \eqref{VillaMass} to our discrete parameter $m$ and we recall the non-negativity property of the Kullback--Leibler divergence and that $\tilde\theta_{1}\neq \tilde\theta_{2}$, we obtain:
We observe that equation \eqref{KL_no_m} is only a function of $\tilde\theta_{1}$ and $\tilde\theta_{2}$ and does not depend on $m$. Thus, $\pi(m)\propto 1$ and, therefore,
\begin{align}
\pi(m)=\dfrac{1}{n-1} \qquad m\in \{1,\ldots, n-1\}.
\label{SimpleOneChgPriorForm}
\end{align}
This prior was used, for instance, in an econometric context by \citet{Koop} with the rationale of giving equal weight to every possible change point location.

\subsection{Multivariate Change Point Problem}
\label{sub_sc_MultipleChgPoints}

In this section, we address the change point problem in its generality by assuming that there are $1\leq k < n$ change points. In particular, for the data $\mathbf{x}^{(n)}=(x_1,\dots,x_n)$, we consider the following sampling distribution 
\begin{equation}\label{SD_multi}
f(\mathbf{x}^{(n)}|\bm{m},\bm{\tilde\theta})=\prod_{i=1}^{m_{1}} f_{1}(x_{i}|\tilde\theta_{1}) \prod_{j=1}^{k-1}\prod_{i=m_{j}+1}^{m_{j+1}} f_{j+1}(x_{i}|\tilde\theta_{j+1})\prod_{i=m_{k}+1}^{n} f_{k+1}(x_{i}|\tilde\theta_{k+1}),
\end{equation}
where $\bm{m}=(m_1,\dots,m_k)$, $1\leq m_1<m_2<\ldots<m_k<n$, is the vector of the change point locations and $\bm{\tilde\theta}=(\tilde\theta_1,\dots,\tilde\theta_k,\tilde\theta_{k+1})$ is the vector of the parameters of the underlying probability distributions. Schematically:
\begin{equation*}
\begin{array}{lcll}
X_{1}&, \ldots, & X_{m_{1}}|\tilde\theta_{1} &\myiid f_{1}(\cdot|\tilde\theta_{1}) \\
X_{m_{1}+1}&, \ldots, &X_{m_2}|\tilde\theta_{2} &\myiid f_{2}(\cdot|\tilde\theta_{2})\\
\vdots&,\ldots, &\vdots &\vdots\ldots\vdots\\
X_{m_{k-1}+1}&, \ldots, &X_{m_k}|\tilde\theta_{k} &\myiid f_{k}(\cdot|\tilde\theta_{k})\\
X_{m_{k}+1}&, \ldots, &X_{n}|\tilde\theta_{k+1} &\myiid f_{k+1}(\cdot|\tilde\theta_{k+1}).\\
\end{array}
\end{equation*}
If $f_1=f_2=\cdots=f_{k+1}$, then it is reasonable to assume that some of the $\theta$'s are different. Without loss of generality, we assume that $\tilde\theta_1\neq\tilde\theta_2\neq\cdots\neq\tilde\theta_k\neq\tilde\theta_{k+1}$. In a similar fashion to the single change point case, we cannot assume $m_k=n$ since we require exactly $k$ change points.

In this case, due to the multivariate nature of the vector $\bm{m}=(m_1,\dots,m_k)$, the derivation of the loss-based prior is not as straightforward as in the one dimensional case. In fact, the derivation of the prior is based on heuristic considerations supported by the below Theorem \ref{lemmaOneChgPoint} (the proof of which is in the Appendix).
In particular, we are able to prove an analogous of equations \eqref{step3case1} and \eqref{step3case2} when only one component is arbitrarily perturbed. 
Let us define the following functions:
\begin{align*}
d^{+1}_j(\bm{\tilde\theta})&=D_{KL}(f_{j+1}(\cdot|\tilde\theta_{j+1})\|f_{j}(\cdot|\tilde\theta_{j}))\\
d^{-1}_j(\bm{\tilde\theta})&=D_{KL}(f_{j}(\cdot|\tilde\theta_{j})\|f_{j+1}(\cdot|\tilde\theta_{j+1})),
\end{align*}
where $j \in \{1, 2, \ldots, k\}$. The following Theorem  is useful to understand the behaviour of the loss-based prior in the general case.\\
\begin{theorem} Let $f(\mathbf{x}^{(n)}|\bm{m},\bm{\tilde\theta})$ be the sampling distribution defined in equation \eqref{SD_multi} and consider  $j\in\lbrace 1,\dots,k\rbrace$. Let $\bm{m}'$ be such that $m'_i=m_i$ for $i\neq j$, and let the component $m'_j$ be such that $m'_j\neq m_j$ and $m_{j-1}<m'_j<m_{j+1}$. Therefore,
$$D_{KL}(f(\mathbf{x}^{(n)}|\bm{m},\bm{\tilde\theta})\|f(\mathbf{x}^{(n)}|\bm{m}',\bm{\tilde\theta})=|m'_j-m_j| d^S_j(\bm{\tilde\theta}),$$
where $S=\emph{sgn}(m'_j-m_j)$.
%$$m_i'=\left\lbrace\begin{array}{ll}
%m_i&\mbox{ if } i\neq j\\
%p& \mbox{ if } i=j
%\end{array}\right.$$
 
\label{lemmaOneChgPoint}
\end{theorem}

Note that, Theorem \ref{lemmaOneChgPoint} states that the minimum Kullback--Leibler divergence is achieved  when $m'_j=m_j+1$ or $m'_j=m_j-1$. This result is not surprising since the Kullback--Leibler divergence measures the degree of similarity between two distributions. The smaller the perturbation caused by changes in one of the parameters is, the smaller the Kullback--Leibler divergence between the two distributions is. Although Theorem \ref{lemmaOneChgPoint} makes a partial statement about the multiple change points scenario, it provides a strong argument for supporting the uniform prior.  Indeed, if now we consider the general case of having $k$ change points, it is straightforward to see that the Kullback--Leibler divergence is minimised when only one of the components of the vector $\bm{m}$ is perturbed by (plus or minus) one unit. As such, the loss-based prior depends on the vector of parameters $\bm{\tilde\theta}$ only, as in the one-dimensional case, yielding the uniform prior for $\bm{m}$.

Therefore, the loss-based prior on the multivariate change point location is  
\begin{align}
\pi(\bm{m})&=\left\lbrace\dbinom{n-1}{k}\right\rbrace^{-1},
\label{MultipleChg}
\end{align}
where $\bm{m}=(m_1,\dots,m_k), 1\leq m_1<m_2<\ldots<m_k<n$.
The denominator in equation \eqref{MultipleChg} has the above form because, for every number of $k$ change points, we are interested in the number of $k$-subsets from a set of $n-1$ elements, which is $\binom{n-1}{k}$. The same prior was also derived in a different way by \citet{Giron}. 

\section{Loss-based Prior on the Number of Change Points}\label{sc_numberCP}

Here, we approach the change point analysis as a model selection problem. In particular, we define a prior on the space of models, where each model represents a certain number of change points (including the case of no change points). The method adopted to define the prior on the space of models is the one introduced in \cite{VillaModel}. 
\begin{figure}[h]
\centering   
\vspace{-6em}
\includegraphics[scale=0.5]{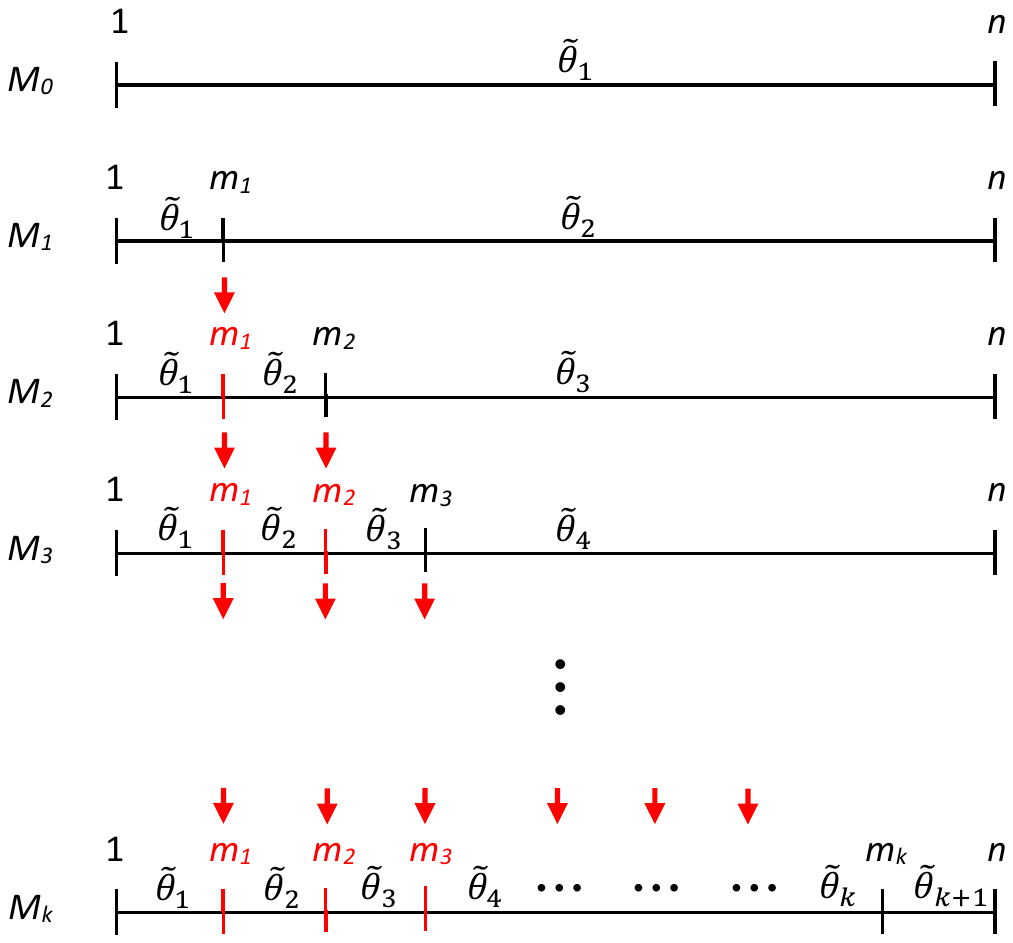}
\vspace{-18em}
\caption{Diagram showing the way we specify our models. The arrows indicate that the respective change point locations remain fixed from the previous model to the current one.}
\label{ModelConstruct}
\end{figure}
We proceed as follows. Assume we have to select from $k+1$ possible models. Let $M_{0}$ be the model with no change points, $M_{1}$ the model with one change point and so on. Generalising, model $M_{k}$ corresponds to the model with $k$ change points. The idea is that the current model encompasses the  change point locations of the previous model. As an example, in model $M_{3}$ the first two change point locations will be the same as in the case of model $M_{2}$. To illustrate the way we envision our models, we have provided Figure \ref{ModelConstruct}. It has to be noted that the construction of the possible models from $M_0$ to $M_k$ can be done in a different way to one here described. Obviously, the approach to define the model priors stays unchanged. Consistently with the notation used in Section 1, 
$$\theta_k=\left\{
\begin{array}{ll}
\tilde{\theta}_1,\ldots, \tilde{\theta}_{k+1},m_1, \ldots, m_k &\mbox{ if } k=1,\dots,n-1\\
\tilde{\theta}_1&\mbox{ if }k=0
\end{array}\right.
$$ represents the vector of parameters of model $M_k$, where $\tilde{\theta}_1,\ldots, \tilde{\theta}_{k+1}$ are the model specific parameters and $m_1, \ldots, m_k$ are the change point locations, as in Figure \ref{ModelConstruct}. 

Based on the way we have specified our models, which are in direct correspondence with the number of change points and their locations, we state Theorem \ref{remark_model} (the proof of which is in the Appendix). 
\vspace*{0.5cm}
\begin{theorem}
Let 
$$D_{KL}(M_{i}\|M_{j})=D_{KL}(f(\mathbf{x}^{(n)}|\theta_i)\|f(\mathbf{x}^{(n)}|\theta_j)).$$
For any $0\leq i<j \leq k$ integers, with $k<n$, and the convention $m_{j+1}=n$, we have the following:
\begin{align*}
 &D_{KL}(M_{i}\|M_{j})=\sum_{q=i+1}^{j}\left[(m_{q+1}-m_{q})\cdot D_{KL}(f_{i+1}(\cdot|\tilde{\theta}_{i+1})\|f_{q+1}(\cdot|\tilde{\theta}_{q+1}))\right],
\end{align*}
and
\begin{align*}
 &D_{KL}(M_{j}\|M_{i})=\sum_{q=i+1}^{j}\left[(m_{q+1}-m_{q})\cdot D_{KL}(f_{q+1}(\cdot|\tilde{\theta}_{q+1})\|f_{i+1}(\cdot|\tilde{\theta}_{i+1}))\right].
\end{align*}
\label{remark_model}
\end{theorem}

The result in Theorem \ref{remark_model} is useful when the model selection exercise is implemented. Indeed, the \cite{VillaModel} approach requires the computation of the Kullback--Leibler divergences in Theorem \ref{remark_model}. Recalling equation \eqref{VillaModelCompactOrg}, the objective model prior probabilities are then given by:
\begin{align}
\Pr(M_{j})\propto &\exp\left\lbrace  \mathbb{E}_{\pi_{j}}\left[\inf_{\theta_{i},i\neq j} D_{KL}(M_j\|M_i)\right]\right\rbrace\qquad j=0, 1, \ldots, k. \label{ModelPriors}
\end{align}

For illustrative purposes, in the Appendix we derive the model prior probabilities to perform model selection among $M_0$, $M_1$ and $M_2$. \\

It is easy to infer from equation \eqref{ModelPriors} that model priors depend on the prior distribution assigned to the model parameters, that is on the level of uncertainty that we have about their true values. For the change point location, a sensible choice is the uniform prior which, as shown in Section \ref{sc_locations}, corresponds to the loss-based prior. For the model specific parameters, we have several options. If one wishes to pursue an objective analysis, intrinsic priors \citep{Intrinsic} may represent a viable solution since they are proper. Nonetheles, the method introduce by \cite{VillaModel} does not require, in principle, an objective choice as long as the priors are proper. Given that we use the latter approach, here we consider subjective priors for the model specific parameters.

{\bf Remark.} In the case where the changes in the underlying sampling distribution are limited to the parameter values, the model prior probabilities defined in \eqref{ModelPriors} follow the uniform distribution. That is, $\Pr(M_j)\propto1$. In the real data example illustrated in Section \ref{sub_sc_British_coal_mine}, we indeed consider a problem where the above case occurs.

\subsection{A special case: selection between $M_0$ and $M_1$}
\label{sub_sc_Model_one_change}
Let us consider the case where we have to estimate whether there is or not a change point in a set of observations. This implies that we have to choose between model $M_0$ (i.e. no change point) and $M_1$ (i.e. one change point). Following our approach, we have:
 
\begin{align}
 \Pr(M_0) \propto \exp\left\{\mathbb{E}_{\pi_{0}}\left[\inf_{\tilde{\theta}_2}D_{KL}(f_{1}(\cdot|\tilde{\theta}_{1})\|f_{2}(\cdot|\tilde{\theta}_{2}))\right]\right\},
 \label{prior_M0_one_chg}
\end{align}
and
\begin{align}
  \Pr(M_1) \propto \exp\left\{\mathbb{E}_{\pi_1}\left[(n-m_{1}) \cdot \inf_{\tilde{\theta}_1}D_{KL}(f_{2}(\cdot|\tilde{\theta}_{2})\|f_{1}(\cdot|\tilde{\theta}_{1}))\right]\right\}
 \label{prior_M1_one_chg}.
\end{align}

Now, let us assume independence between the prior on the change point location  and the prior on the parameters of the underlying sampling distributions, that is $\pi_1(m_1,\tilde{\theta}_1, \tilde{\theta}_2)=\pi_1(m_1) \pi_1(\tilde{\theta}_1, \tilde{\theta}_2)$. Let us further recall that, as per equation \eqref{MultipleChg}, $\pi_1(m_1)=1/(n-1)$.  As such, we observe that the model prior probability on $M_1$ becomes:
 \begin{align}
  \Pr(M_1) \propto \exp\left\lbrace\left(\dfrac{n}{2}\right)\mathbb{E}_{\pi_1(\tilde{\theta}_1, \tilde{\theta}_2)}\left[\inf_{\tilde{\theta}_1}D_{KL}(f_{2}(\cdot|\tilde{\theta}_{2})\|f_{1}(\cdot|\tilde{\theta}_{1}))\right]\right\rbrace.
 \label{prior_M1_trans_one_chg}
\end{align}

We notice that the model prior probability for model $M_1$ is increasing when the sample size increases. This behaviour occurs whether there is or not a change point in the data. We propose to address the above problem by using a non-uniform prior for $m_1$. A reasonable alternative, which works quite well in practice, would be the following shifted binomial as prior:
 \begin{align}
  \pi_1(m_1)=\dbinom{n-2}{m_1-1}\left(\dfrac{n-1}{n}\right)^{m_1-1}\left(\dfrac{1}{n}\right)^{n-m_1-1}, 1\leq m_1 \leq n-1.
  \label{Pseudo_binom}
 \end{align}
To argument the choice of \eqref{Pseudo_binom}, we note that, as $n$ increases, the probability mass will be more and more concentrated towards the upper end of the support. Therefore, from equations \eqref{prior_M1_one_chg} and \eqref{Pseudo_binom} follows:
 \begin{align}
  \Pr(M_1) \propto \exp\left\lbrace\left(\dfrac{2n-2}{n}\right)\mathbb{E}_{\pi_1(\tilde{\theta}_1, \tilde{\theta}_2)}\left[\inf_{\tilde{\theta}_1}D_{KL}(f_{2}(\cdot|\tilde{\theta}_{2})\|f_{1}(\cdot|\tilde{\theta}_{1}))\right]\right\rbrace.
 \label{prior_M1_pseudo_binom}
\end{align}
For the more general case where we consider more than two models, the problem highlighted in equation \eqref{prior_M1_trans_one_chg} vanishes. 

\section{Change Point Analysis on Simulated Data}\label{sc_simulation}
In this section, we present the results of several simulation studies based on the methodologies discussed in Sections \ref{sc_locations} and \ref{sc_numberCP}. We start with a scenario involving discrete distributions in the context of the one change point problem. We then show the results obtained when we consider continuous distributions for the case of two change points. The choice of the underlying sampling distributions is in line with \citet{VillaModel}.

\subsection{Single sample}

\paragraph*{Scenario 1.} The first scenario concerns the choice between models $M_0$ and $M_1$. Specifically, for $M_0$ we have:
$$X_{1}, X_{2}, \ldots, X_{n}|p \myiid \mbox{Geometric}(p)$$
and for $M_1$ we have:
\begin{align*}
X_{1}, X_{2}, \ldots, X_{m_{1}}&|p \myiid \mbox{Geometric}(p) \\
X_{m_{1}+1}, X_{m_{1}+2}, \ldots, X_{n}&|\lambda \myiid \mbox{Poisson}(\lambda)
\end{align*}

Let us denote with $f_1(\cdot|p)$ and $f_2(\cdot|\lambda)$ the probability mass functions of the Geometric and the Poisson distributions, respectively. The priors for the parameters of $f_1$ and $f_2$ are $
p \sim \mbox{Beta}(a,b)$ and $\lambda \sim \mbox{Gamma}(c,d)$.

In the first simulation, we sample $n=100$ observations from model $M_0$ with $p=0.8$. To perform the change point analysis, we have chosen the following parameters for the priors on $p$ and $\lambda$: $a=2$, $b=2$, $c=3$ and $d=1$. Applying the approach introduced in Section \ref{sc_numberCP}, we obtain $\Pr(M_0)\propto 1.59$ and $\Pr(M_1)\propto 1.81$. These model priors yield the posterior distribution probabilities (refer to equation \eqref{modelposterior}) $\Pr(M_0|\mathbf{x}^{(n)})=0.92$ and $\Pr(M_1|\mathbf{x}^{(n)})=0.08$. As expected, the selection process strongly indicates the true model as $M_0$. Table \ref{Discrete_Results_1} reports the above probabilities including other information, such as the appropriate Bayes factors.

\begin{figure}[h]
\centering
\includegraphics[scale=0.6]{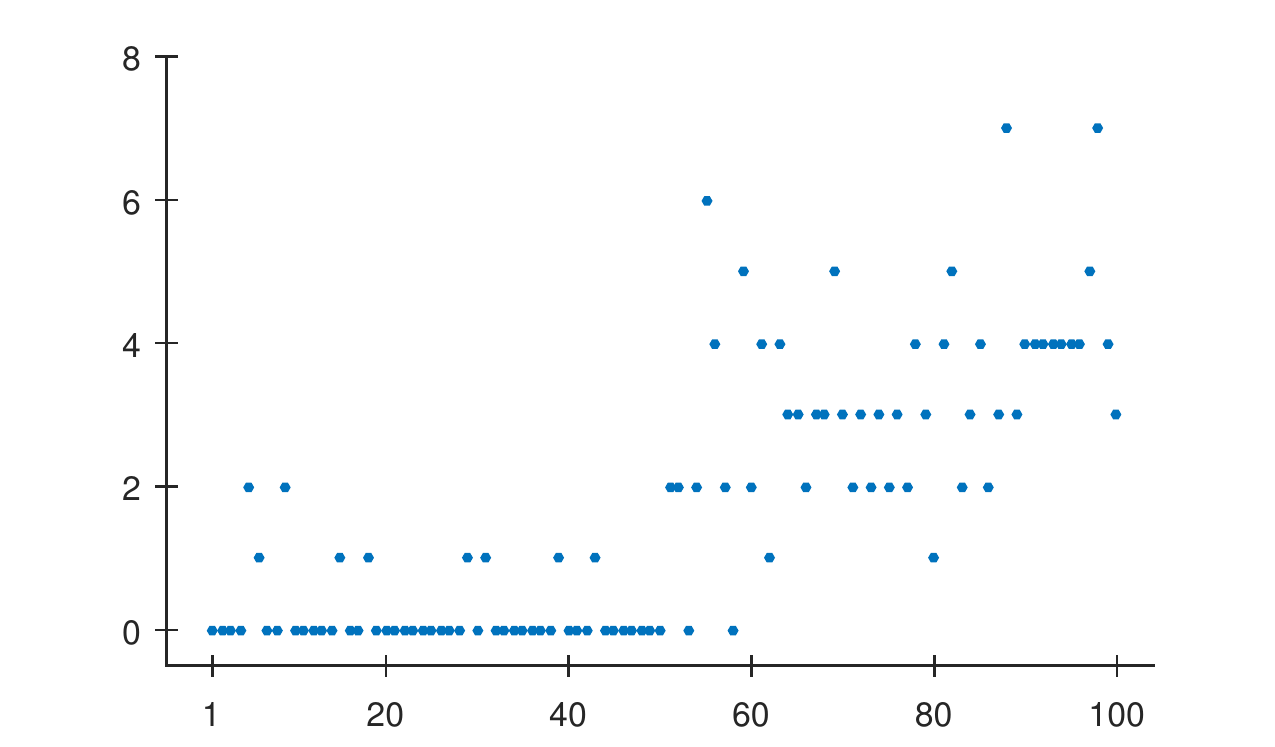}
\caption{Scatter plot of the data simulated from model $M_1$ in Scenario 1.}
\label{fig_GP}
\end{figure}

The second simulation looked at the opposite setup, that is we sample $n=100$ observations from $M_1$, with $p=0.8$ and $\lambda=3$. We have sampled 50 data points from the Geometric distribution and the remaining 50 data points from the Poisson distribution. In Figure \ref{fig_GP}, we have plotted the simulated sample, where it is legitimate to assume a change in the underlying distribution. Using the same prior parameters as above, we obtain $\Pr(M_0|\mathbf{x}^{(n)})=0.06$ and $\Pr(M_1|\mathbf{x}^{(n)})=0.94$. Again, the model selection process is assigning heavy posterior mass to the true model $M_1$. These results are further detailed in Table \ref{Discrete_Results_1}.

\begin{table}[h]
\centering
\begin{tabular}{c|cc}
\hline 
• & \multicolumn{2}{c}{True model} \\ 
%\hline 
• & $M_0$ & $M_1$ \\ 
\hline 
$\Pr(M_0)$ & 0.47 & 0.47 \\ 
%\hline 
$\Pr(M_1)$ & 0.53 & 0.53 \\ 
%\hline 
$B_{01}$ & 12.39 & 0.08 \\ 
%\hline 
$B_{10}$ & 0.08 & 12.80 \\ 
%\hline 
$\Pr(M_0|\mathbf{x}^{(n)})$ & 0.92 & 0.06 \\ 
%\hline 
$\Pr(M_1|\mathbf{x}^{(n)})$ & 0.08 & 0.94 \\ 
\hline 
\end{tabular}
\caption{Model prior, Bayes factor and model posterior probabilities for the change point analysis in Scenario 1. We considered samples from, respectively, model $M_0$ and model $M_1$.}
\label{Discrete_Results_1}
\end{table}

\paragraph*{Scenario 2.} In this scenario we consider the case where we have to select among three models, that is model $M_0$:
\begin{align}
X_{1}, X_{2}, \ldots, X_{n}|\lambda, \kappa &\myiid \mbox{Weibull}(\lambda, \kappa) ,\label{Simulation_M_0}
\end{align}
model $M_1$:
\begin{align}
X_{1}, X_{2}, \ldots, X_{m_{1}}|\lambda, \kappa &\myiid \mbox{Weibull}(\lambda, \kappa)\nonumber \\
X_{m_{1}+1}, X_{m_{1}+2}, \ldots, X_{n}|\mu, \tau &\myiid \mbox{Log-normal}(\mu, \tau) \label{Simulation_M_1},
\end{align}
with $1\leq m_{1}\leq n-1$ being the location of the single change point, and model $M_2$:
\begin{align}
X_{1}, X_{2}, \ldots, X_{m_{1}}| \lambda, \kappa &\myiid \mbox{Weibull}(\lambda, \kappa)\nonumber\\
X_{m_{1}+1}, X_{m_{1}+2}, \ldots, X_{m_{2}}| \mu, \tau &\myiid \mbox{Log-normal}(\mu, \tau)\nonumber\\
X_{m_{2}+1}, X_{m_{2}+2}, \ldots, X_{n}|\alpha, \beta &\myiid \mbox{Gamma}(\alpha, \beta), \label{Simulation_M_2}
\end{align}
with $1\leq m_{1}<m_{2}\leq n-1$ representing the locations of the two change points, such that $m_1$ corresponds exactly to the same location as in model $M_1$. Analogously to the previous scenario, we sample from each model in turn and perform the selection to detect the number of change points.

Let $f_{1}(\cdot|\lambda, \kappa), f_{2}(\cdot|\mu, \tau)$ and $f_{3}(\cdot|\alpha, \beta)$ represent the Weibull, Log-normal and Gamma densities, respectively, with $\tilde\theta_1=(\lambda, \kappa)$, $\tilde\theta_2=(\mu, \tau)$ and $\tilde\theta_3=(\alpha, \beta)$. We assume a Normal prior on $\mu$ and Gamma priors on all the other parameters as follows:
$$\lambda\sim\mbox{Gamma}(1.5,1) \qquad \kappa\sim\mbox{Gamma}(5,1) \qquad \mu\sim\mbox{Normal}(0.05,1),$$
$$\quad\tau\sim\mbox{Gamma}(16,1) \qquad \alpha\sim\mbox{Gamma}(10,1) \qquad \beta\sim\mbox{Gamma}(0.2,0.1).$$

\begin{figure}[H]
\centering
\includegraphics[scale=0.6]{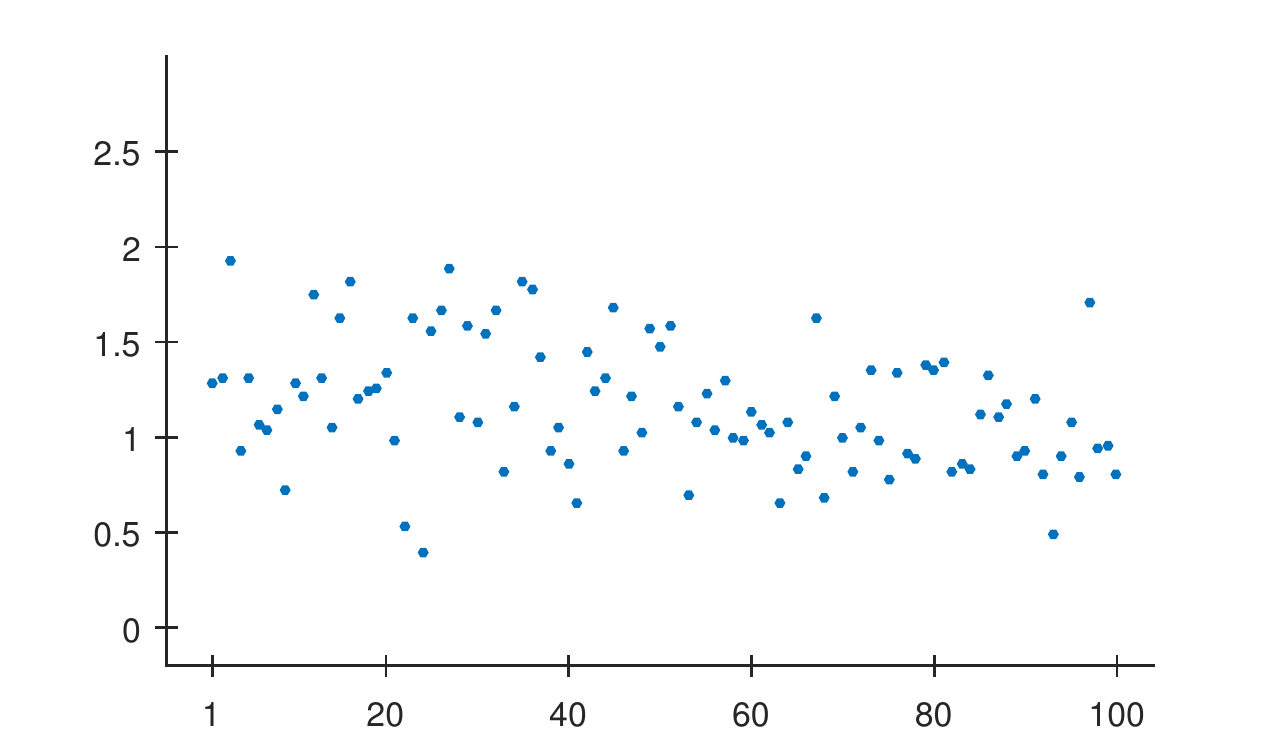}
\caption{Scatter plot of the observations simulated from model $M_1$ in Scenario 2.}
\label{One_change}
\end{figure}

In the first exercise, we have simulated $n=100$ observations from model $M_0$, where we have set $\lambda=1.5$ and $\kappa=5$.  We obtain the following model priors: $\Pr(M_0)\propto1.09$, $\Pr(M_1)\propto1.60$ and $\Pr(M_2)\propto1.37$, yielding the posteriors $\Pr(M_0|\mathbf{x}^{(n)})=0.96$, $\Pr(M_1|\mathbf{x}^{(n)})=0.04$ and $\Pr(M_2|\mathbf{x}^{(n)})=0.00$. We then see that the approach assigns high mass to the true model $M_0$. Table \ref{TableResults} reports the above probabilities and the corresponding Bayes factors.
\begin{table}[h]
\centering
\begin{tabular}{c|ccc}
\hline 
• & \multicolumn{3}{c}{True model} \\ 
%\hline 
• & $M_0$ & $M_1$ & $M_2$ \\ 
\hline 
$\Pr(M_0)$ & 0.27 & 0.27 & 0.27 \\ 
%\hline 
$\Pr(M_1)$ & 0.39 & 0.39 & 0.39 \\ 
%\hline 
$\Pr(M_2)$ & 0.34 & 0.34 & 0.34 \\
%hline
$B_{01}$ & 36.55 & $3.24\times10^{-4}$ & $4.65\times10^{-40}$ \\ 
%\hline 
$B_{02}$ & $1.84\times 10^3$ & 0.02 & $1.27\times10^{-45}$ \\ 
%\hline 
$B_{12}$ & 50.44 & 55 & $2.72\times10^{-6}$ \\
%\hline
$\Pr(M_0|\mathbf{x}^{(n)})$ & 0.96 & 0.00 & 0.00\\ 
%\hline 
$\Pr(M_1|\mathbf{x}^{(n)})$ & 0.04 & 0.98 & 0.00 \\
%\hline 
$\Pr(M_2|\mathbf{x}^{(n)})$ & 0.00 & 0.02 & 1.00 \\
\hline 
\end{tabular}
\caption{Model prior, Bayes factor and model posterior probabilities for the change point analysis in Scenario 2. We considered samples from, respectively, model $M_0$, model $M_1$ and model $M_2$.}
\label{TableResults}
\end{table}
The second simulation was performed by sampling 50 observations from a Weibull with parameter values as in the previous exercise, and the remaining 50 observations from a Log-normal density with location parameter $\mu=0.05$ and scale parameter $\tau=16$. The data is displayed in Figure \ref{One_change}.
The model posterior probabilities are $\Pr(M_0|\mathbf{x}^{(n)})=0.00$, $\Pr(M_1|\mathbf{x}^{(n)})=0.98$ and $\Pr(M_2|\mathbf{x}^{(n)})=0.02$, which are reported in Table \ref{TableResults}. In this case as well, we see that the model selection procedure indicates $M_1$ as the true model, as expected.
\begin{figure}[H]
\centering
\includegraphics[scale=0.6]{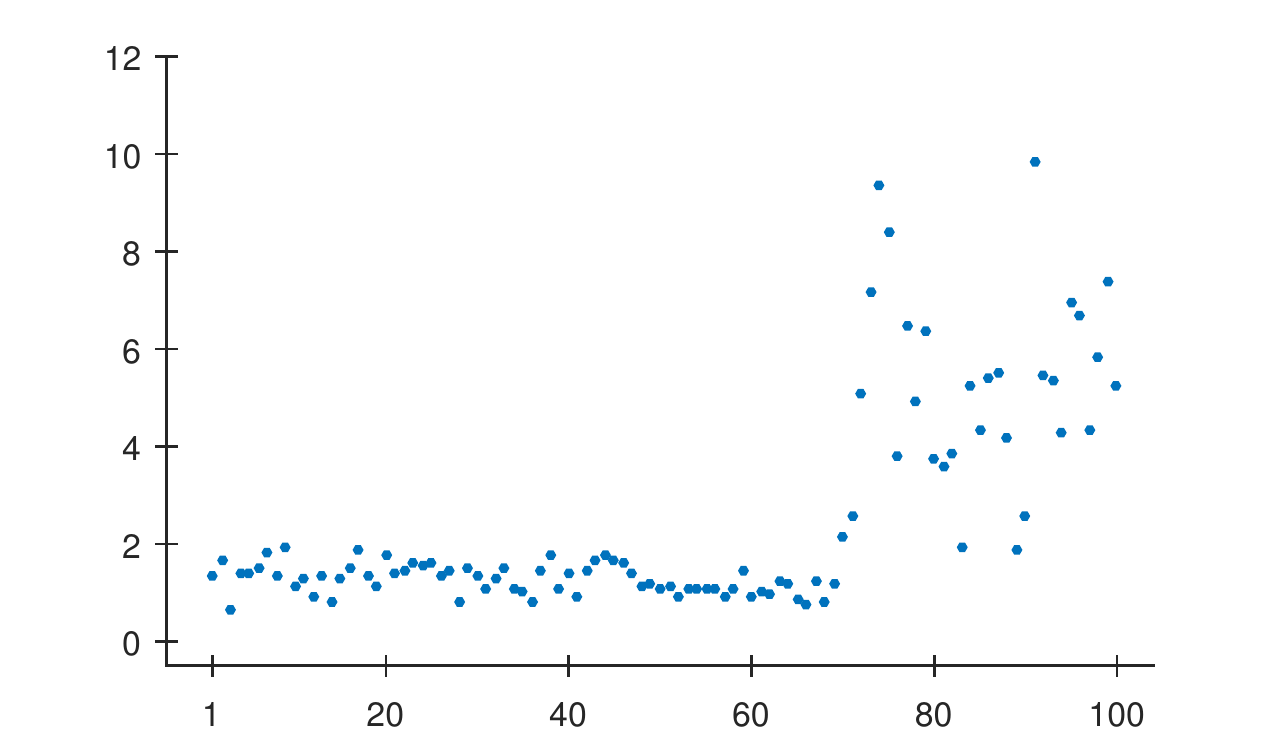}
\caption{Scatter plot of the observations simulated from model $M_2$ in Scenario 2.}
\label{Two_change}
\end{figure}
Finally, for the third simulation exercise we sample 50 and 20 data points from, respectively, a Weibull and a Log-normal with parameter values as defined above, and the last 30 observations are sampled from a Gamma distribution with parameters $\alpha=10$ and $\beta=2$. From Table \ref{TableResults}, we note that the posterior distribution on the model space accumulates on the true model $M_2$.

 \subsection{Frequentist Analysis}
In this section, we perform a frequentist analysis of the performance of the proposed prior by drawing repeated samples from different scenarios. In particular, we look at a two change points problem where the sampling distributions are Student-$t$ with different degrees of freedom. In this scenario, we perform the analysis with $60$ repeated samples generated by different densities with the same mean values.

Then, we repeat the analysis of Scenario 2 by selecting $100$ samples for $n=500$ and $n=1500$. We consider different sampling distributions with the same mean and variance. In this scenario, where we added the further constraint of the equal variance, it is interesting to note that the change in distribution is captured when we increase the sample size, meaning that we learn more about the true sampling distributions. 

%It is not unusual in practice to have relatively large data sets (e.g. financial data). As such, the above is not a limitation of the approach but, actually, a point of strength.

We also compare the performances of the loss-based prior with the uniform prior when we analyse the scenario with different sampling distributions. Namely, Weibull/Log-normal/Gamma.  It is interesting to note that the uniform prior is unable to capture the change in distribution even for a large sample size. On the contrary, the loss-based prior is able to detect the number of change points when $n=1500$. Furthermore, for $n=500$, even though both priors are not able to detect the change points most of the times, the loss-based prior has a higher frequency of success when compared to the uniform prior. 

\paragraph*{Scenario 3.}
In this scenario, we consider the case where the sampling distributions belong to the same family, that is Student-$t$, where the true model has two change points. In particular, let $f_{1}(\cdot|\nu_1), f_{2}(\cdot|\nu_2)$ and $f_{3}(\cdot|\nu_3)$ represent the densities of three standard $t$ distributions, respectively. We assume that $\nu_1, \nu_2$ and $\nu_3$ are positive integers strictly greater than one so to have defined mean for each density. Note that this allows us to compare distributions of the same family with equal mean. The priors assigned to the number of degrees of freedom assume a parameter space of positive integers strictly larger than 1. As such, we define them as follows:
$$\nu_1\sim 2+\mbox{Poisson}(30) \qquad  \nu_2\sim 2+\mbox{Poisson}(3) \qquad \nu_3\sim 2+\mbox{Poisson}(8).$$
In this experiment, we consider 60 repeated samples, each of size $n=300$ and with the following structure: 
\begin{itemize}
\item $X_1, \dots, X_{100}$ from a Student-$t$ distribution with $\nu_1=30$,
\item $X_{101}, \dots, X_{200}$ from a Student-$t$ distribution with $\nu_2=3$,
\item $X_{201}, \dots, X_{300}$ from a Student-$t$ distribution with $\nu_3=8$.
\end{itemize}
Table \ref{TableResults-T-distrib} reports the frequentist results of the simulation study. First, note that $P(M_1)=P(M_2)=P(M_3)=1/3$ as per the Remark in Section 3. For all the simulated samples, the loss-based prior yields a posterior with the highest probability assigned to the true model $M_2$. We also note that the above posterior is on average $0.75$  with a variance $0.02$, making the inferential procedure extremely accurate. 
\begin{table}[h]
\centering
\begin{tabular}{c|c|c|c}
\hline 
• & Mean posterior & Variance posterior &  \pbox{15cm}{\relax\ifvmode\centering\fi \rule{0pt}{1em} Freq. true model} \\ 
%\hline 
\hline 
%$\Pr(M_0)$ & 0.33 [$3.37\times10^{-9}$]&   \\ 
%%\hline 
%$\Pr(M_1)$ & 0.33 [$3.59\times10^{-9}$]&   \\ 
%%\hline 
%$\Pr(M_2)$ & 0.34 [$1.39\times10^{-8}$] &   \\
%hline
%$B_{01}$ & 36.55 & $3.24\times10^{-4}$ & $4.65\times10^{-40}$ \\ 
%\hline 
%$B_{02}$ & $1.84\times 10^3$ & 0.02 & $1.27\times10^{-45}$ \\ 
%\hline 
%$B_{12}$ & 50.44 & 55 & $2.72\times10^{-6}$ \\
%\hline
$\Pr(M_0|\mathbf{x}^{(n)})$ & 0.01 & $3.84\times10^{-4}$&  0/60\\ 
%\hline 
$\Pr(M_1|\mathbf{x}^{(n)})$ & 0.24 & 0.0160  & 0/60 \\
%\hline 
$\Pr(M_2|\mathbf{x}^{(n)})$ & 0.75 & 0.0190 & 60/60 \\
\hline 
%\pbox{15cm}{\relax\ifvmode\centering\fi \rule{0pt}{1em} Frequency of identifying \\ the respective model} && \\
%\hline
\end{tabular}
\caption{Average model posterior probabilities, variance and frequency of true model for the Scenario 3 simulation exercise.}
\label{TableResults-T-distrib}
\end{table}

\paragraph*{Scenario 4.}

In this scenario, we perform repeated sampling from the setup described in scenario 2 above, where the true model has two change points. In particular, we draw $100$ samples with $n=500$ and $n=1500$. For $n=500$, the loss-based prior probabilities are $P(M_0)=0.18$, $P(M_1)=0.16$ and $P(M_2)=0.66$. For $n=1500$, the loss-based prior probabilities are $P(M_0)=0.015$, $P(M_1)=0.014$ and $P(M_2)=0.971$. The simulation results are reported, respectively, in Table \ref{TableResults-WLG-Conjugate-500} and in Table \ref{TableResults-WLG-Conjugate-1500}.  The two change point locations for $n=500$ are at the $171$st and $341$st observations. For $n=1500$, the first change point is the $501$st observation, while the second is at the $1001$st observation. We note that there is a sensible improvement in detecting the true model, using the loss-based prior, when the sample size increases. In particular, we move from $30\%$ to $96\%$.

\begin{table}[H]
\centering
\begin{tabular}{c|c|c|c}
\hline 
• & Mean posterior & Variance posterior &  \pbox{15cm}{\relax\ifvmode\centering\fi \rule{0pt}{1em} Freq. true model} \\
%\hline 
\hline 
%$\Pr(M_0)$ & 0.18 [$6.66\times10^{-14}$]&   \\ 
%%\hline 
%$\Pr(M_1)$ & 0.16 [$1.53\times10^{-12}$]&   \\ 
%%\hline 
%$\Pr(M_2)$ & 0.66 [$9.54\times10^{-13}$] &   \\
%hline
%$B_{01}$ & 36.55 & $3.24\times10^{-4}$ & $4.65\times10^{-40}$ \\ 
%\hline 
%$B_{02}$ & $1.84\times 10^3$ & 0.02 & $1.27\times10^{-45}$ \\ 
%\hline 
%$B_{12}$ & 50.44 & 55 & $2.72\times10^{-6}$ \\
%\hline
$\Pr(M_0|\mathbf{x}^{(n)})$ & $9.88\times10^{-4}$​ &$2.60\times10^{-5}$&  0/100\\ 
%\hline 
$\Pr(M_1|\mathbf{x}^{(n)})$ & 0.63 & 0.0749 & 70/100 \\
%\hline 
$\Pr(M_2|\mathbf{x}^{(n)})$ & 0.37 & 0.0745& 30/100 \\
\hline 
%\pbox{15cm}{\relax\ifvmode\centering\fi \rule{0pt}{1em} Frequency of identifying \\ the respective model} && \\
%\hline
\end{tabular}
\caption{Average model posterior probabilities, variance and frequency of true model for the Scenario 4 simulation exercise with $n=500$ and the loss-based prior.}
\label{TableResults-WLG-Conjugate-500}
\end{table}

\begin{table}[h]
\centering
\begin{tabular}{c|c|c|c}
\hline 
• & Mean posterior & Variance posterior &  \pbox{15cm}{\relax\ifvmode\centering\fi \rule{0pt}{1em} Freq. true model} \\
%\hline 
\hline 
%$\Pr(M_0)$ & 0.015 [$3.46\times10^{-18}$]&   \\ 
%%\hline 
%$\Pr(M_1)$ & 0.014 [$1.51\times10^{-14}$]&   \\ 
%%\hline 
%$\Pr(M_2)$ & 0.971 [$1.46\times10^{-14}$] &   \\
%hline
%$B_{01}$ & 36.55 & $3.24\times10^{-4}$ & $4.65\times10^{-40}$ \\ 
%\hline 
%$B_{02}$ & $1.84\times 10^3$ & 0.02 & $1.27\times10^{-45}$ \\ 
%\hline 
%$B_{12}$ & 50.44 & 55 & $2.72\times10^{-6}$ \\
%\hline
$\Pr(M_0|\mathbf{x}^{(n)})$ & $1.33\times10^{-13}​$ &$1.76\times10^{-24}$&  0/100\\ 
%\hline 
$\Pr(M_1|\mathbf{x}^{(n)})$ & 0.08 &0.0200  & 4/100 \\
%\hline 
$\Pr(M_2|\mathbf{x}^{(n)})$ & 0.92 &0.0200 & 96/100 \\
\hline 
%\pbox{15cm}{\relax\ifvmode\centering\fi \rule{0pt}{1em} Frequency of identifying \\ the respective model} && \\
%\hline
\end{tabular}
\caption{Average model posterior probabilities, variance and frequency of true model for the Scenario 4 simulation exercise with $n=1500$ and the loss-based prior. }
\label{TableResults-WLG-Conjugate-1500}
\end{table}

To compare the loss-based prior with the uniform prior we have run the simulation on the same data samples used above. The results for $n=500$ and $n=1500$ are in Table \ref{Unif_TableResults-WLG-Conjugate-500} and in Table \ref{Unif_TableResults-WLG-Conjugate-1500}, respectively. Although we can observe an improvement when the sample size increases, the uniform prior does not lead to a clear detection of the true model for both sample sizes.

\begin{table}[H]
\centering
\begin{tabular}{c|c|c|c}
\hline 
• & Mean posterior & Variance posterior &  \pbox{15cm}{\relax\ifvmode\centering\fi \rule{0pt}{1em} Freq. true model} \\
%\hline 
\hline 
%hline
%$B_{01}$ & 36.55 & $3.24\times10^{-4}$ & $4.65\times10^{-40}$ \\ 
%\hline 
%$B_{02}$ & $1.84\times 10^3$ & 0.02 & $1.27\times10^{-45}$ \\ 
%\hline 
%$B_{12}$ & 50.44 & 55 & $2.72\times10^{-6}$ \\
%\hline
$\Pr(M_0|\mathbf{x}^{(n)})$ & $16\times 10^{-4}$& $7.15\times10^{-5}$ &  0/100\\ 
%\hline 
$\Pr(M_1|\mathbf{x}^{(n)})$ & 0.82 & 0.0447 & 91/100 \\
%\hline 
$\Pr(M_2|\mathbf{x}^{(n)})$ & 0.18 & 0.0443 & 9/100 \\
\hline 
%\pbox{15cm}{\relax\ifvmode\centering\fi \rule{0pt}{1em} Frequency of identifying \\ the respective model} && \\
%\hline
\end{tabular}
\caption{Average model posterior probabilities, variance and frequency of true model for the Scenario 4 simulation exercise with $n=500$ and the uniform prior.}
\label{Unif_TableResults-WLG-Conjugate-500}
\end{table}

\begin{table}[h]
\centering
\begin{tabular}{c|c|c|c}
\hline 
• & Mean posterior & Variance posterior &  \pbox{15cm}{\relax\ifvmode\centering\fi \rule{0pt}{1em} Freq. true model} \\
%\hline 
\hline 
%$\Pr(M_0)$ & 0.3333 &   \\ 
%%\hline 
%$\Pr(M_1)$ & 0.3333 &   \\ 
%%\hline 
%$\Pr(M_2)$ & 0.3333 &   \\
%hline
%$B_{01}$ & 36.55 & $3.24\times10^{-4}$ & $4.65\times10^{-40}$ \\ 
%\hline 
%$B_{02}$ & $1.84\times 10^3$ & 0.02 & $1.27\times10^{-45}$ \\ 
%\hline 
%$B_{12}$ & 50.44 & 55 & $2.72\times10^{-6}$ \\
%\hline
$\Pr(M_0|\mathbf{x}^{(n)})$ & $8.64\times10^{-12}$ & $7.45\times10^{-21}$&  0/100\\ 
%\hline 
$\Pr(M_1|\mathbf{x}^{(n)})$ & 0.501 &0.1356  & 49/100 \\
%\hline 
$\Pr(M_2|\mathbf{x}^{(n)})$ & 0.499 & 0.1356 & 51/100 \\
\hline 
%\pbox{15cm}{\relax\ifvmode\centering\fi \rule{0pt}{1em} Frequency of identifying \\ the respective model} && \\
%\hline
\end{tabular}
\caption{Average model posterior probabilities, variance and frequency of true model for the Scenario 4 simulation exercise with $n=1500$ and the uniform prior.}
\label{Unif_TableResults-WLG-Conjugate-1500}
\end{table}

\begin{figure}[H]
\centering
\includegraphics[scale=0.7]{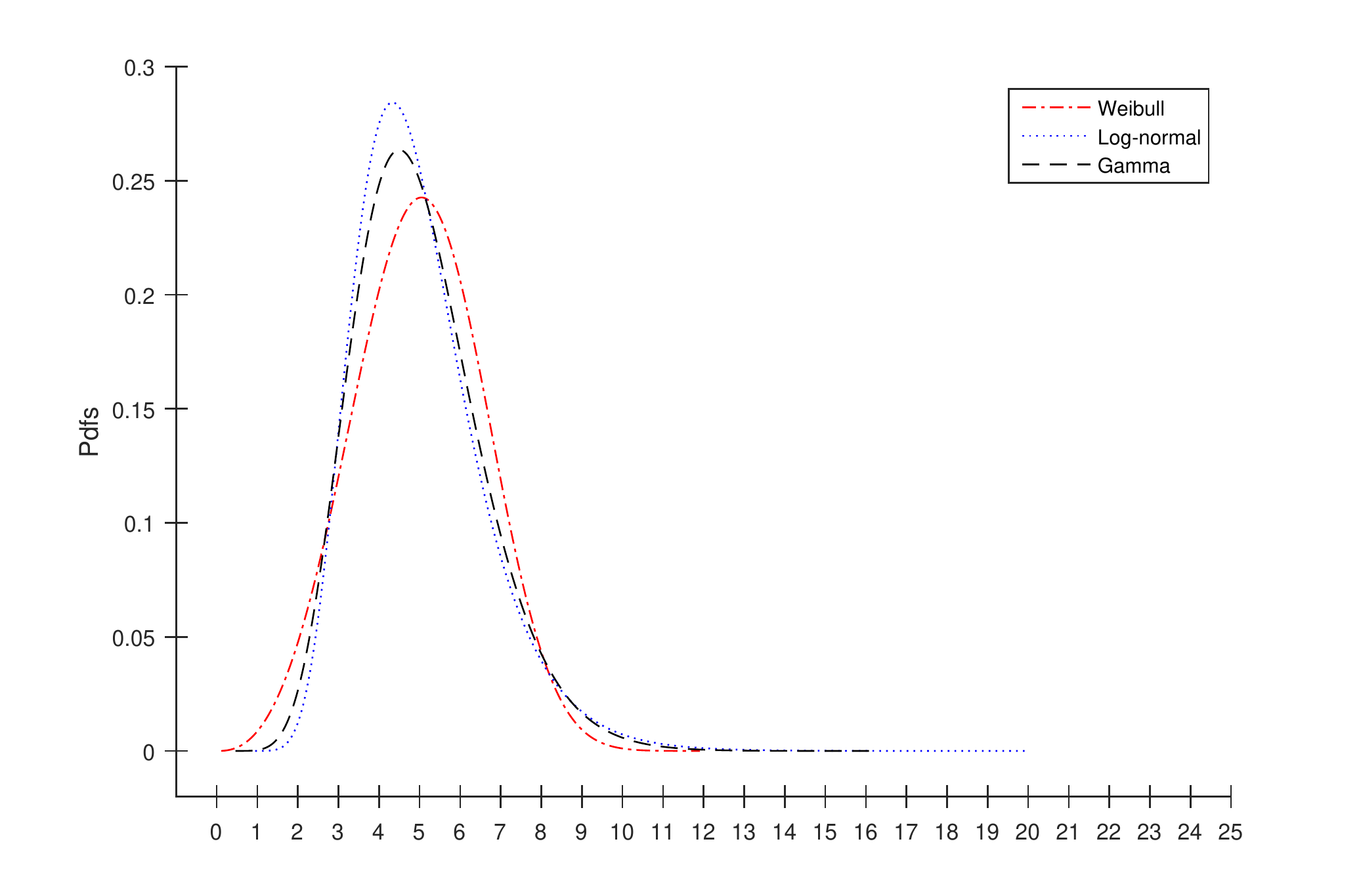}
\caption{The densities of Weibull($\lambda,\kappa$), Log-normal($\mu,\tau$) and  Gamma($\alpha,\beta$) with the same mean (equal to 5) and the same variance (equal to 2.5).}
%\caption{The density functions of Weibull($\lambda,\kappa$), Log-normal($\mu,\tau$) and  Gamma($\alpha,\beta$) which were plotted using the \texttt{\textbf{R}} function \texttt{density()} where the bandwidth was set to 40 times the default bandwidth}
\label{fig_PDFs_WLG}
\end{figure}

Finally, we conclude this section with a remark. One may wonder why the change point detection requires an increasing in the sample size, and the reply can be inferred from Figure \ref{fig_PDFs_WLG}, which displays the density functions of the distributions employed in this scenario. As it can be observed, the densities are quite similar, which is not surprising since these distributions have the same means and the same variances. The above similarity can also be appreciated in terms of Hellinger distance, see Table \ref{Hellinger_TableResults}. In other words, from Figure \ref{fig_PDFs_WLG} we can see that the main differences in the underlying distributions are in the tail areas. It is therefore necessary to have a relatively large number of observations in order to be able to discern differences in the densities, because in this case only we would have a sufficient representation of the whole distribution.

\begin{table}[H]
\centering
\begin{tabular}{c|ccc}
\hline 
• & \multicolumn{3}{c}{Hellinger distances} \\ 
%\hline 
• & Weibull($\lambda,\kappa$) & Log-normal($\mu,\tau$) & Gamma($\alpha,\beta$) \\ 
\hline 
Weibull($\lambda,\kappa$) &  & 0.1411996  & 0.09718282 \\ 
%\hline 
Log-normal($\mu,\tau$) &  &  & 0.04899711 \\ 
\hline 
%Gamma($\alpha,\beta$) &  &  & 0.34 \\
%\hline 
\end{tabular}
\caption{Hellinger distances between all the pairs formed from a Weibull($\lambda,\kappa$), Log-normal($\mu,\tau$) and  Gamma($\alpha,\beta$). The six hyperparameters are such that the distributions have the same mean=5 and same variance=2.5.}
\label{Hellinger_TableResults}
\end{table}

\section{Change Point Analysis on Real Data}\label{sc_realdata}
In this section, we illustrate the proposed approach applied to real data. We first consider a well known data set which has been extensively studied in the literature of the change point analysis, that is the British coal-mining disaster data \citep{Carlin}. The second set of data we consider refers to the daily returns of the S\&P 500 index observed over a period of four years. The former data set will be investigated in Section \ref{sub_sc_British_coal_mine}, while the latter in Section \ref{sub_sc_Financial_data}.

\subsection{British Coal-Mining Disaster Data}
\label{sub_sc_British_coal_mine}
\begin{figure}[h]
\centering   
\includegraphics[scale=0.6]{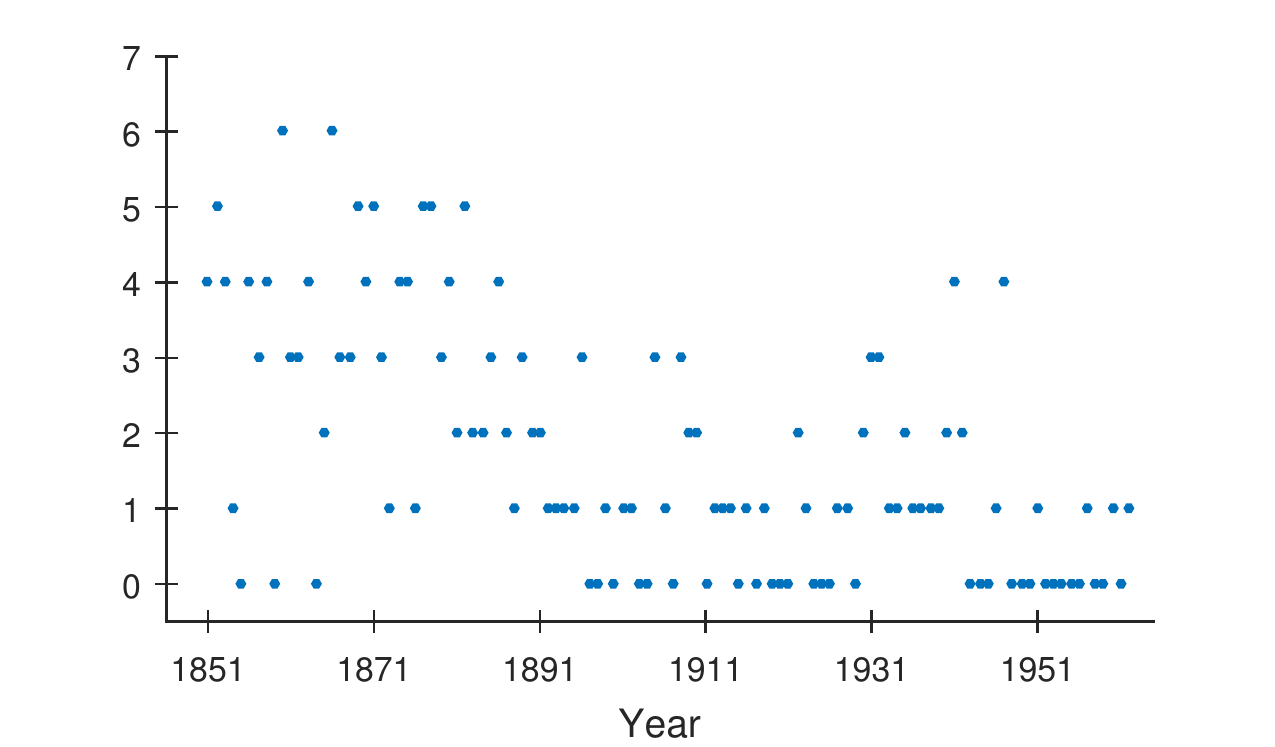}
\caption{Scatter plot of the British coal-mining disaster data.}
\label{British_coalmine}
\end{figure}

The British coal-mining disaster data consists of the yearly number of deaths for the British coal miners over the period 1851-1962. It is believed that the change in the working conditions, and in particular, the enhancement of the security measures, led to a decrease in the number of deaths. This calls for a model which can take into account a change in the underlying distribution around a certain observed year. With the proposed methodology we wish to detect if the assumption is appropriate. In particular, if a model with one change point is more suitable to represent the data than a model where no changes in the sampling distribution are assumed. Figure \ref{British_coalmine} shows the number of deaths per year in the British coal-mining industry from 1851 to 1962.
As in \citet{Chib}, we assume a Poisson sampling distribution with a possible change in the parameter value. That is
\begin{align}\label{m1_mining}
X_{1}, X_{2}, \ldots, X_{m}|\phi_{1} &\myiid \mbox{Poisson}(\phi_{1})\nonumber \\
X_{m+1}, X_{m+2}, \ldots, X_{n}|\phi_{2} &\myiid \mbox{Poisson}(\phi_{2}),
\end{align}
where $m$ is the unknown location of the single change point, such that $1\leq m \leq n$, and a $\mbox{Gamma}(2,1)$ is assumed for $\phi_1$ and $\phi_2$. The case $m=n$ corresponds to the scenario with no change point, that is model $M_0$. The case $m<n$ assumes one change point, that is model $M_1$.

Let $f_1(\cdot|\phi_1)$ and $f_2(\cdot|\phi_2)$ be the Poisson distributions with parameters $\phi_1$ and $\phi_2$, respectively. Then, the analysis is performed by selecting between model $M_0$, that is when the sampling distribution is $f_1$, and model $M_1$, where the sampling distribution is $f_1$ up to a certain $m<n$ and $f_2$ from $m+1$ to $n$.

As highlighted in the Remark at the end of Section \ref{sc_numberCP}, the prior on the model space is the discrete uniform distribution, that is $\Pr(M_0)=\Pr(M_1)=0.5$. The proposed model selection approach leads to the Bayes factors $B_{01}=1.61 \times 10^{-13}$ and $B_{10}=6.20 \times 10^{12}$, where it is obvious that the odds are strongly in favour of model $M_1$. Indeed, we have $\Pr(M_1|{\bf x}^{(n)})\approx1$.

\subsection{Daily S\&P 500 Absolute Log-Return Data}
\label{sub_sc_Financial_data}
\begin{figure}[h]
\centering
\includegraphics[scale=0.6]{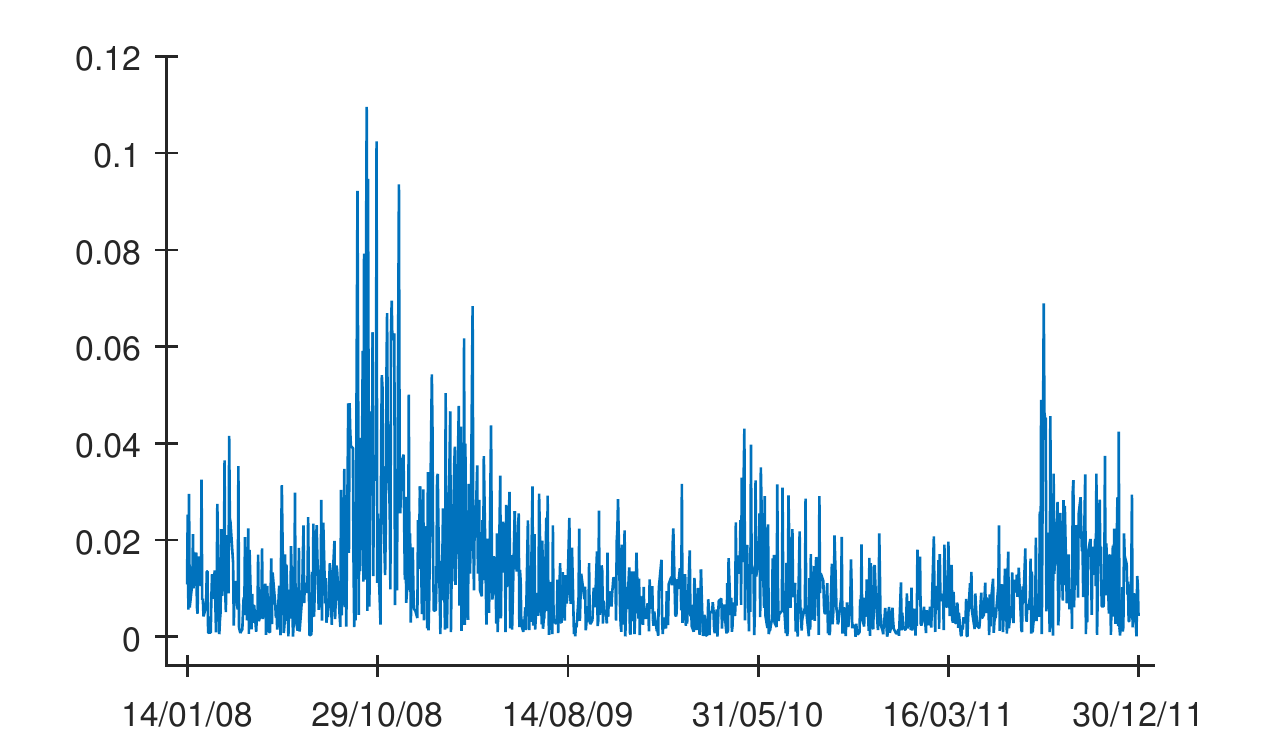}
\caption{Absolute daily log-returns of the S\&P500 index from 14/01/08 to 30/12/11.}
\label{fig_sp500}
\end{figure}

The second real data analysis aims to detect change points in the absolute value of the daily logarithmic returns of the S\&P500 index observed from the 14/01/2008 to the 31/12/2011 (see Figure \ref{fig_sp500}). As underlying sampling distributions we consider the Weibull and the Log-normal \citep{Yu}, and the models among which we select are as follows. $M_0$ is a $\mbox{Weibull}(\lambda,\kappa)$, $M_1$ is formed by a $\mbox{Weibull}(\lambda,\kappa)$ and a $\mbox{Log-normal}(\mu_1,\tau_1)$ and, finally, $M_2$ is formed by a $\mbox{Weibull}(\lambda,\kappa)$, a $\mbox{Log-normal}(\mu_1,\tau_1)$ and a $\mbox{Log-normal}(\mu_2,\tau_2)$. An interesting particularity of this problem is that we will consider a scenario where the changes are in the underlying distribution or in the parameter values of the same distribution. As suggested in Section 4.1.3 of \cite{KassRaftery}, due to the large sample size of the data set, we could approximate the Bayes factor by using the Schwarz criterion. Therefore, in this case the specification of the priors for the parameters of the underlying distributions is not necessary.
From the results in Table \ref{Finance_TableResults}, we see that the model indicated by the proposed approach is $M_2$. In other words, there is very strong indication that there are two change points in the data set.
\begin{table}[h]
\centering
\begin{tabular}{c|ccc}
%\hline 
%\hline
\hline 
$\Pr(M_0)$ & 0.36 \\ 
%\hline 
$\Pr(M_1)$ & 0.32 \\ 
%\hline 
$\Pr(M_2)$ & 0.32 \\
%hline
$B_{01}$ & $7.72\times10^{18}$ \\ 
%\hline 
$B_{02}$ & $3.30\times10^{-3}$ \\ 
%\hline 
$B_{12}$ & $4.28\times10^{-22}$ \\
%\hline
$\Pr(M_0|\mathbf{x}^{(n)})$ & 0.00\\ 
%\hline 
$\Pr(M_1|\mathbf{x}^{(n)})$ & 0.00 \\
%\hline 
$\Pr(M_2|\mathbf{x}^{(n)})$ & 1.00 \\
\hline 
\end{tabular}
\caption{Model prior, Bayes factor and model posterior probabilities for the S\&P500 change point analysis.}
\label{Finance_TableResults}
\end{table}
From Table \ref{Finance_TableResults}, we note that the prior on model $M_1$ and $M_2$ assigned by the proposed method are the same. This is not surprising as the only difference between the two models is an additional Log-normal distribution with different parameter values. 
\section{Conclusion}\label{Conclusion}
Bayesian inference in change point problems under the assumption of minimal prior information has not been deeply explored in the past, as the limited literature on the matter shows.

We contribute to the area by deriving an objective prior distribution to detect change point locations, when the number of change points is known a priori. As a change point location can be interpreted as a discrete parameter, we apply recent results in the literature \citep{VillaMass} to make inference. The resulting prior distribution, which is the discrete uniform distribution, it is not new in the literature \citep{Giron}, and therefore can be considered as a validation of the proposed approach.

A second major contribution is in defining an objective prior on the number of change points, which has been approached by considering the problem as a model selection exercise. The results of the proposed method on both simulated and real data, show the strength of the approach in estimating the number of change points in a series of observations. A point to note is the generality of the scenarios considered. Indeed, we consider situations where the change is in the value of the parameter(s) of the underlying sampling distribution, or in the distribution itself. Of particular interest is the last real data analysis (S\&P 500 index), where we consider a scenario where we have both types of changes, that is the distribution for the first change point and on the parameters of the distribution for the second.

The aim of this work was to set up a novel approach to address change point problems. In particular, we have selected prior densities for the parameters of the models to reflect a scenario of equal knowledge, in the sense that model priors are close to represent a uniform distribution. Two remarks are necessary here. First, in the case prior information about the true value of the parameters is available, and one wishes to exploit it, the prior densities will need to reflect it and, obviously, the model prior will be impacted by the choice. Second, in applications it is recommended that some sensitivity analysis is performed, so to investigate if and how the choice of the parameter densities affects the selection process.

\section*{Acknowledgements}
Fabrizio Leisen was supported by the European Community's Seventh Framework Programme [FP7/2007-2013] under grant agreement no: 630677. Cristiano Villa was supported by the  Royal Society Research Grant no: RG150786.

\newpage

\section*{Appendix}
\setcounter{subsection}{0}
\renewcommand{\thesubsection}{\Alph{subsection}}
\subsection{Model prior probabilities to select among models $M_0$, $M_1$ and $M_2$}
Here, we show how model prior probabilities can be derived for the relatively simple case of selecting among scenarios with no change points ($M_0$), one change point ($M_1$) or two change points ($M_2$). First, by applying the result in Theorem \ref{remark_model}, we derive the Kullback--Leibler divergences between any two models. That is:
\begin{itemize}
\item the prior probability for model $M_{0}$ depends on the following quantities:
\begin{align*}
 D_{KL}(M_0\|M_1)=&(n-m_{1})\cdot D_{KL}(f_{1}(\cdot|\tilde{\theta}_{1})\|f_{2}(\cdot|\tilde{\theta}_{2}))\\
  D_{KL}(M_0\|M_2)=&(m_{2}-m_{1})\cdot D_{KL}(f_{1}(\cdot|\tilde{\theta}_{1})\|f_{2}(\cdot|\tilde{\theta}_{2}))\\
&+(n-m_{2})\cdot D_{KL}(f_{1}(\cdot|\tilde{\theta}_{1})\|f_{3}(\cdot|\tilde{\theta}_{3}))
\end{align*}

\item the prior probability for model $M_{1}$ depends on the following quantities:
\begin{align*}
 D_{KL}(M_{1}\|M_{2})=&(n-m_{2})\cdot D_{KL}(f_{2}(\cdot|\tilde{\theta}_{2})\|f_{3}(\cdot|\tilde{\theta}_{3}))\\
 D_{KL}(M_1\|M_0)=&(n-m_{1})\cdot D_{KL}(f_{2}(\cdot|\tilde{\theta}_{2})\|f_{1}(\cdot|\tilde{\theta}_{1}))
\end{align*}

\item the prior probability for model $M_{2}$ depends on the following quantities:
\begin{align*}
 D_{KL}(M_2\|M_1)=&(n-m_{2})\cdot D_{KL}(f_{3}(\cdot|\tilde{\theta}_{3})\|f_{2}(\cdot|\tilde{\theta}_{2}))\\
 D_{KL}(M_2\|M_0)=&(m_{2}-m_{1})\cdot D_{KL}(f_{2}(\cdot|\tilde{\theta}_{2})\|f_{1}(\cdot|\tilde{\theta}_{1}))\\
&+(n-m_{2})\cdot D_{KL}(f_{3}(\cdot|\tilde{\theta}_{3})\|f_{1}(\cdot|\tilde{\theta}_{1}))
\end{align*}
\end{itemize}
%Recalling the need for $\displaystyle\inf_{\theta_{i},i\neq j} D_{KL}(M_j\|M_i)$, we obtain:
The next step is to derive the minimum Kullback--Leibler divergence computed at each model:
\begin{itemize}
\item for model $M_{0}$:
\begin{align*}
 \inf_{\theta_1}D_{KL}(M_0\|M_1)=&\underbrace{\left[\inf_{m_{1}\neq n}(n-m_{1})\right]}_{1}\cdot \left[\inf_{\tilde{\theta}_2}D_{KL}(f_{1}(\cdot|\tilde{\theta}_{1})\|f_{2}(\cdot|\tilde{\theta}_{2}))\right]\\=& \inf_{\tilde{\theta}_2}D_{KL}(f_{1}(\cdot|\tilde{\theta}_{1})\|f_{2}(\cdot|\tilde{\theta}_{2}))\\
   \inf_{\theta_2}D_{KL}(M_0\|M_2)=&\underbrace{\left[\inf_{m_{1}\neq m_{2}}(m_{2}-m_{1})\right]}_{1}\cdot \left[\inf_{\tilde{\theta}_{2}}D_{KL}(f_{1}(\cdot|\tilde{\theta}_{1})\|f_{2}(\cdot|\tilde{\theta}_{2}))\right]\\
&+\underbrace{\left[\inf_{m_{2} \neq n}(n-m_{2})\right]}_{1}\cdot \left[\inf_{\tilde{\theta}_3}D_{KL}(f_{1}(\cdot|\tilde{\theta}_{1})\|f_{3}(\cdot|\tilde{\theta}_{3}))\right]\\ =& \inf_{\tilde{\theta}_{2}}D_{KL}(f_{1}(\cdot|\tilde{\theta}_{1})\|f_{2}(\cdot|\tilde{\theta}_{2}))+\inf_{\tilde{\theta}_3}D_{KL}(f_{1}(\cdot|\tilde{\theta}_{1})\|f_{3}(\cdot|\tilde{\theta}_{3}))
\end{align*}

\item for model $M_{1}$:
\begin{align*}
 \inf_{\theta_2}D_{KL}(M_{1}\|M_{2})=&\underbrace{\left[\inf_{m_2 \neq n}(n-m_{2})\right]}_{1}\cdot \left[\inf_{\tilde{\theta}_3}D_{KL}(f_{2}(\cdot|\tilde{\theta}_{2})\|f_{3}(\cdot|\tilde{\theta}_{3}))\right]\\=&\inf_{\tilde{\theta}_3}D_{KL}(f_{2}(\cdot|\tilde{\theta}_{2})\|f_{3}(\cdot|\tilde{\theta}_{3}))\\
 \inf_{\theta_0=\tilde{\theta}_1}D_{KL}(M_1\|M_0)=&(n-m_{1})\cdot \inf_{\tilde{\theta}_1}D_{KL}(f_{2}(\cdot|\tilde{\theta}_{2})\|f_{1}(\cdot|\tilde{\theta}_{1}))
\end{align*}

\item for model $M_{2}$:
\begin{align*}
 \inf_{\theta_1}D_{KL}(M_2\|M_1)=&(n-m_{2})\cdot \inf_{\tilde{\theta}_2}D_{KL}(f_{3}(\cdot|\tilde{\theta}_{3})\|f_{2}(\cdot|\tilde{\theta}_{2}))\\
 \inf_{\theta_0=\tilde{\theta}_1}D_{KL}(M_2\|M_0)=&(m_{2}-m_{1})\cdot \inf_{\tilde{\theta}_1}D_{KL}(f_{2}(\cdot|\tilde{\theta}_{2})\|f_{1}(\cdot|\tilde{\theta}_{1}))\\
&+(n-m_{2})\cdot \inf_{\tilde{\theta}_1}D_{KL}(f_{3}(\cdot|\tilde{\theta}_{3})\|f_{1}(\cdot|\tilde{\theta}_{1}))
\end{align*}
\end{itemize}
Therefore, the model prior probabilities can be computed through equation \eqref{ModelPriors}, so that:
\begin{itemize}
\item the model prior probability $\Pr(M_{0})$ is proportional to the exponential of the minimum between:
\begin{align*}
 &\left\lbrace\mathbb{E}_{\pi_{0}}\left[\inf_{\tilde{\theta}_2}D_{KL}(f_{1}(\cdot|\tilde{\theta}_{1})\|f_{2}(\cdot|\tilde{\theta}_{2}))\right],\mathbb{E}_{\pi_{0}}\left[\inf_{\tilde{\theta}_{2}}D_{KL}(f_{1}(\cdot|\tilde{\theta}_{1})\|f_{2}(\cdot|\tilde{\theta}_{2}))\right.\right. \nonumber \\ &\left.\left. \hspace*{0.3cm}+\inf_{\tilde{\theta}_3}D_{KL}(f_{1}(\cdot|\tilde{\theta}_{1})\|f_{3}(\cdot|\tilde{\theta}_{3}))\right]\right\rbrace
\end{align*}

\item the model prior probability $\Pr(M_{1})$ is proportional to the exponential of the minimum between:
\begin{align*}
 &\left\lbrace\mathbb{E}_{\pi_{1}}\left[\inf_{\tilde{\theta}_3}D_{KL}(f_{2}(\cdot|\tilde{\theta}_{2})\|f_{3}(\cdot|\tilde{\theta}_{3}))\right], \right. \nonumber \\ &\left. \hspace*{0.3cm} \mathbb{E}_{\pi_1}\left[(n-m_{1}) \cdot \inf_{\tilde{\theta}_1}D_{KL}(f_{2}(\cdot|\tilde{\theta}_{2})\|f_{1}(\cdot|\tilde{\theta}_{1}))\right]\right\rbrace
\end{align*}

\item the model prior probability $\Pr(M_{2})$ is proportional to the exponential of the minimum between:
\begin{align*}
  &\left\lbrace\mathbb{E}_{\pi_{2}}\left[(n-m_{2})\cdot \inf_{\tilde{\theta}_2}D_{KL}(f_{3}(\cdot|\tilde{\theta}_{3})\|f_{2}(\cdot|\tilde{\theta}_{2}))\right],\right.\nonumber\\&\left.\hspace*{0.3cm}\mathbb{E}_{\pi_{2}}\left[(m_{2}-m_{1})\cdot \inf_{\tilde{\theta}_1}D_{KL}(f_{2}(\cdot|\tilde{\theta}_{2})\|f_{1}(\cdot|\tilde{\theta}_{1}))+(n-m_{2})\right.\right.\nonumber\\&\left.\left.\hspace*{0.3cm}\cdot\inf_{\tilde{\theta}_1}D_{KL}(f_{3}(\cdot|\tilde{\theta}_{3})\|f_{1}(\cdot|\tilde{\theta}_{1}))\right]\right\rbrace
\end{align*}
\end{itemize}
\newpage
\subsection{Proofs}
\subsubsection*{Proof of Theorem \ref{lemmaOneChgPoint}}
%\begin{proof}
We distinguish two cases: $S=+1$ and $S=-1$. When $S=+1$, equivalent to $m_j<m_j^{\prime}$:
\begin{align}
&D_{KL}(f(\mathbf{x}^{(n)}|\bm{m},\bm{\tilde\theta})\|f(\mathbf{x}^{(n)}|\bm{m}',\bm{\tilde\theta}))=\int f(\mathbf{x}^{(n)}|\bm{m},\bm{\tilde\theta}) \cdot \ln\left(\dfrac{f(\mathbf{x}^{(n)}|\bm{m},\bm{\tilde\theta})}{f(\mathbf{x}^{(n)}|\bm{m}',\bm{\tilde\theta})}\right)\,\mathrm{d}\mathbf{x}^{(n)} \nonumber \\
=&\int  f(\mathbf{x}^{(n)}|\bm{m},\bm{\tilde\theta}) \cdot \left[\sum_{i=m_{j}+1}^{m_{j}^{\prime}} \ln\left(\dfrac{f_{j+1}(x_{i}|\tilde\theta_{j+1})}{f_{j}(x_{i}|\tilde\theta_{j})}\right)\right]\,\mathrm{d}\mathbf{x}^{(n)} \nonumber \\
=&\mathlarger{\sum_{i=m_{j}+1}^{m_{j}^{\prime}}} \int f(\mathbf{x}^{(n)}|\bm{m},\bm{\tilde\theta}) \cdot \left[\ln\left(\dfrac{f_{j+1}(x_{i}|\tilde\theta_{j+1})}{f_{j}(x_{i}|\tilde\theta_{j})}\right)\right]\,\mathrm{d}\mathbf{x}^{(n)} \nonumber \\
=&\mathlarger{\sum_{i=m_{j}+1}^{m_{j}^{\prime}}} \left\lbrace 1^{n-1} \cdot \int f_{j+1}(x_{i}|\tilde\theta_{j+1}) \cdot \left[\ln\left(\dfrac{f_{j+1}(x_{i}|\tilde\theta_{j+1})}{f_{j}(x_{i}|\tilde\theta_{j})}\right)\right]\,\mathrm{d}x_{i}\right\rbrace \nonumber \\
=&\sum_{i=m_{j}+1}^{m_{j}^{\prime}} D_{KL}(f_{j+1}(x_{i}|\tilde\theta_{j+1})\|f_{j}(x_{i}|\tilde\theta_{j})) \nonumber \\
=&(m_j^{\prime}-m_{j}) \cdot D_{KL}(f_{j+1}(\cdot|\tilde\theta_{j+1})\|f_{j}(\cdot|\tilde\theta_{j})) \nonumber \\
=&(m_j^{\prime}-m_{j}) \cdot d_j^{+1}(\bm{\tilde\theta}).
\label{demopositive}
\end{align}
When $S=-1$, equivalent to $m_j>m_j^{\prime}$, in a similar fashion, we get 
\begin{align}
&D_{KL}(f(\mathbf{x}^{(n)}|\bm{m},\bm{\tilde\theta})\|f(\mathbf{x}^{(n)}|\bm{m}',\bm{\tilde\theta}))
=(m_{j}-m_{j}^{\prime}) \cdot d_j^{-1}(\bm{\tilde\theta})
\label{demonegative}
\end{align}
From equations \eqref{demopositive} and \eqref{demonegative}, we get the result in Theorem \ref{lemmaOneChgPoint}.
%\end{proof}

\subsubsection*{Proof of Theorem \ref{remark_model}}
%\begin{proof}
We recall that the model parameter $\theta_i$ is the vector $(m_1, m_2, \ldots, m_i, \tilde{\theta}_1, \tilde{\theta}_2, \ldots, \tilde{\theta}_{i+1})$, where $i=0, 1, \ldots, k$. Here, $\tilde{\theta}_1, \tilde{\theta}_2,\ldots, \tilde{\theta}_{i+1}$ represent the parameters of the underlying sampling distributions considered under model $M_i$ and $m_1, m_2, \ldots, m_i$ are the respective $i$ change point locations. In this setting, 
\begin{align}
f(\mathbf{x}^{(n)}|\theta_i)=\prod_{r=1}^{m_{1}} f_{1}(x_{r}|\tilde{\theta}_{1})\prod_{t=1}^{i-1}\prod_{r=m_{t}+1}^{m_{t+1}} f_{t+1}(x_{r}|\tilde{\theta}_{t+1})\prod_{r=m_{i}+1}^{n} f_{i+1}(x_{r}|\tilde{\theta}_{i+1})
\label{proof2_0}
\end{align}

We proceed to the computation of $D_{KL}(M_{i}\|M_{j})$, that is the Kullback--Leibler divergence  introduced in Section \ref{sc_numberCP}. Similarly to the proof of Theorem \ref{lemmaOneChgPoint}, we obtain the following result.  
\begin{align*}
D_{KL}(M_{i}\|M_{j})=&\sum_{r=m_{i+1}+1}^{m_{i+2}} \int f(\mathbf{x}^{(n)}|\theta_i)\ln\left(\dfrac{f_{i+1}(x_{r}|\tilde\theta_{i+1})}{f_{i+2}(x_{r}|\tilde\theta_{i+2})}\right)\,\mathrm{d}\mathbf{x}^{(n)}\nonumber \\
&+ \sum_{r=m_{i+2}+1}^{m_{i+3}} \int f(\mathbf{x}^{(n)}|\theta_i)\ln\left(\dfrac{f_{i+1}(x_{r}|\tilde\theta_{i+1})}{f_{i+3}(x_{r}|\tilde\theta_{i+3})}\right)\,\mathrm{d}\mathbf{x}^{(n)}+\nonumber \\
&\ldots+ \sum_{r=m_{j}+1}^{n} \int f(\mathbf{x}^{(n)}|\theta_i) \ln\left(\dfrac{f_{i+1}(x_{r}|\tilde\theta_{i+1})}{f_{j+1}(x_{r}|\tilde\theta_{j+1})}\right)\,\mathrm{d}\mathbf{x}^{(n)}.
\end{align*}
Given equation \eqref{proof2_0}, if we integrate out the variables not involved in the logarithms, we obtain
\begin{align*}
 D_{KL}(M_{i}\|M_{j})=&(m_{i+2}-m_{i+1})\cdot D_{KL}(f_{i+1}(\cdot|\tilde{\theta}_{i+1})\|f_{i+2}(\cdot|\tilde{\theta}_{i+2})) \\
&+(m_{i+3}-m_{i+2})\cdot D_{KL}(f_{i+1}(\cdot|\tilde{\theta}_{i+1})\|f_{i+3}(\cdot|\tilde{\theta}_{i+3}))+\\
&\ldots+(n-m_{j})\cdot D_{KL}(f_{i+1}(\cdot|\tilde{\theta}_{i+1})\|f_{j+1}(\cdot|\tilde{\theta}_{j+1})).
\end{align*}
In a similar fashion, it can be shown that
\begin{align*}
 D_{KL}(M_{j}\|M_{i})=&(m_{i+2}-m_{i+1})\cdot D_{KL}(f_{i+2}(\cdot|\tilde{\theta}_{i+2})\|f_{i+1}(\cdot|\tilde{\theta}_{i+1}))\\
&+(m_{i+3}-m_{i+2})\cdot D_{KL}(f_{i+3}(\cdot|\tilde{\theta}_{i+3})\|f_{i+1}(\cdot|\tilde{\theta}_{i+1}))+\\
&\ldots+(n-m_{j})\cdot D_{KL}(f_{j+1}(\cdot|\tilde{\theta}_{j+1})\|f_{i+1}(\cdot|\tilde{\theta}_{i+1}))
\end{align*}
\appendix

\end{document}